\documentclass[10pt,letterpaper,english,10pt,final,journal,letterpaper,twocolumn,twoside]{IEEEtran}
\usepackage[T1]{fontenc}
\usepackage[latin9]{inputenc}
\usepackage{color}
\usepackage{babel}
\usepackage{float}
\usepackage{mathrsfs}
\usepackage{amsmath}
\usepackage{amsthm}
\usepackage{amssymb}
\usepackage{stmaryrd}
\usepackage{graphicx}
\usepackage{esint}
\usepackage[unicode=true,pdfusetitle,
 bookmarks=true,bookmarksnumbered=false,bookmarksopen=false,
 breaklinks=false,pdfborder={0 0 0},pdfborderstyle={},backref=false,colorlinks=true]
 {hyperref}

\makeatletter

\input{mysymbol.sty}

\floatstyle{ruled}
\newfloat{algorithm}{tbp}{loa}
\providecommand{\algorithmname}{Algorithm}
\floatname{algorithm}{\protect\algorithmname}

\theoremstyle{plain}
\newtheorem{thm}{\protect\theoremname}
\theoremstyle{plain}
\newtheorem{lem}[thm]{\protect\lemmaname}
\theoremstyle{plain}
\newtheorem{prop}[thm]{\protect\propositionname}
\theoremstyle{remark}
\newtheorem{rem}[thm]{\protect\remarkname}
\theoremstyle{definition}
\newtheorem{defn}[thm]{\protect\definitionname}
\theoremstyle{plain}
\newtheorem{cor}[thm]{\protect\corollaryname}

\IEEEoverridecommandlockouts

\usepackage{dsfont}
\usepackage{babel}
\usepackage{relsize}
\usepackage{subdepth}
\allowdisplaybreaks

\author{
\IEEEauthorblockN{Dionysios S. Kalogerias, \textit{Member, IEEE}, Mark Eisen, \textit{Member, IEEE}
       \\George J. Pappas, \textit{Fellow, IEEE}, and Alejandro Ribeiro}
\vspace{-13pt}
\thanks{D. S. Kalogerias, G. J. Pappas, and A. Ribeiro are with the Dept. of Electrical \& Systems Engineering, University of Pennsylvania, Philadelphia, PA 19104, USA (e-mail: \{dionysis, pappasg, aribeiro\}@seas.upenn.edu). M. Eisen is with Intel Corporation, Hillsboro, OR 97124, USA (e-mail: mark.eisen@intel.com). This work is supported by the Intel Science and Technology Center for Wireless Autonomous Systems and NSF under Grant CPS 1837253.}}

\usepackage{dsfont}

\usepackage{babel}

\usepackage{relsize}
\usepackage{subdepth}

\usepackage{babel}
\providecommand{\definitionname}{Definition}
\providecommand{\lemmaname}{Lemma}
\providecommand{\corollaryname}{Corollary}
\providecommand{\theoremname}{Theorem}

\usepackage{babel}
\providecommand{\definitionname}{Definition}

\providecommand{\lemmaname}{Lemma}
\providecommand{\remarkname}{Remark}
\providecommand{\theoremname}{Theorem}

\usepackage{url}

\DeclareFontFamily{OT1}{pzc}{}
\DeclareFontShape{OT1}{pzc}{m}{it}{<-> s * [1.200] pzcmi7t}{}
\DeclareMathAlphabet{\mathpzc}{OT1}{pzc}{m}{it}

\usepackage[noadjust]{cite}
\usepackage{filecontents}

\usepackage{changepage}
\usepackage{stfloats}

\newtheorem{assumption}{Assumption}

\def\Halmos{$\blacksquare$}

\definecolor{ballblue}{rgb}{0.13, 0.67, 0.8}
\hypersetup{citecolor=ballblue}

\usepackage{enumitem}

\@ifundefined{showcaptionsetup}{}{%
 \PassOptionsToPackage{caption=false}{subfig}}
\usepackage{subfig}

\makeatother

\providecommand{\corollaryname}{Corollary}
\providecommand{\definitionname}{Definition}
\providecommand{\lemmaname}{Lemma}
\providecommand{\propositionname}{Proposition}
\providecommand{\remarkname}{Remark}
\providecommand{\theoremname}{Theorem}

\begin{document}
\title{Model-Free Learning of Optimal\\
Ergodic Policies in Wireless Systems}
\maketitle
\begin{abstract}
Learning optimal resource allocation policies in wireless systems can be effectively achieved by formulating finite dimensional constrained programs which depend on system configuration, as well as the adopted learning parameterization. 
The interest here is in cases where system models are unavailable, prompting methods that probe the wireless system with candidate policies, and then use observed performance to determine better policies. This generic procedure is difficult because of the need to cull accurate gradient estimates out of these limited system queries. This paper constructs and exploits smoothed surrogates of constrained ergodic resource allocation problems, the gradients of the former being representable exactly as averages of finite differences that can be obtained through limited system probing. Leveraging this unique property, we develop a new model-free primal-dual algorithm for learning optimal ergodic resource allocations, while we rigorously analyze the relationships between original policy search problems and their surrogates, in both primal and dual domains.
First, we show that both primal and dual domain surrogates are uniformly consistent approximations of their corresponding original finite dimensional counterparts. Upon further assuming the use of near-universal policy parameterizations, we also develop explicit bounds on the gap between optimal values of initial, infinite dimensional resource allocation problems, and dual values of their parameterized smoothed surrogates. In fact, we show that this duality gap decreases at a linear rate relative to smoothing and universality parameters. Thus, it can be made arbitrarily small at will, also justifying our proposed primal-dual algorithmic recipe. Numerical simulations confirm the effectiveness of our approach.
\end{abstract}

\begin{IEEEkeywords}
Wireless Systems, Stochastic Resource Allocation, Zeroth-order Optimization,
Constrained Nonconvex Optimization, Deep Learning, Lagrangian Duality,
Strong Duality.
\end{IEEEkeywords}

\setlength{\textfloatsep}{5pt} 

\vspace{-4pt}
\section{\label{sec:Optimal-Resource-Allocation}Introduction and Problem Formulation}

We investigate optimal wireless communication systems operating over realizations of random fading channels $\bbH\in\ccalH\subseteq\reals^{N_{\bbH}}$ with distribution $\ccalM_{\bbH}$. Resources such as transmission power and channel access are allocated to jointly maximize the service levels of one or multiple users, in a certain sense. Due to randomness of $\bbH$, a reasonable objective is to optimize quality of service in an ergodic regime, i.e., by averaging all possible instantaneous service levels relative to the fading distribution ${\cal M}_{\bbH}$. Then, optimal wireless system design
may be abstracted to a stylized base resource allocation problem of the form \cite{Ribeiro2012}
\renewcommand{\arraystretch}{1.2}
\begin{equation}
\begin{array}{rl}
\underset{\bbx,\bbp(\cdot)}{\mathrm{maximize}} & g^{o}(\bbx)\\
\mathrm{subject\,to} & \bbx\le\mathbb{E}\{\bbf(\bbp(\bbH),\bbH)\}\\
& \bbg(\bbx)\ge{\bf 0},\,(\bbx,\bbp)\in{\cal X}\times{\cal P}
\end{array}.\label{eq:Base_Problem}
\end{equation}
%
In \eqref{eq:Base_Problem}, the policy $\bbp:\ccalH\to\bbR^{N_{R}}$ maps fading states $\bbH$ to $N_R$ resource allocation decisions $\bbp(\bbH)$, the function $\bbf:\bbR^{N_{R}}\times\ccalH\to\bbR^{N_{S}}$ maps decisions and fading values to $N_S$ \textit{instantaneous} service level metrics, the average of which bounds the ergodic metrics $\bbx\in\bbR^{N_S}$, whose worth we evaluate through the utilities $g^{o}:\bbR^{N_{S}}\to\bbR$ and $\bbg:\bbR^{N_{S}}\to\bbR^{N_{\bbg}}$. Ergodic performances are further restricted to the set $\ccalX\subseteq\bbR^{N_{S}}$ and resource allocations are further restricted to the set $\ccalP$, the latter inducing \textit{pointwise} constraints on each individual value $\bbp(\bbH)$ of every candidate policy $\bbp$ \cite{Ribeiro2012}, for each fading realization $\bbH$.

Problem \eqref{eq:Base_Problem} conveniently abstracts several resource allocation tasks of practical importance. It is relatively straightforward to see that particular cases of \eqref{eq:Base_Problem} appear naturally in, e.g., point-to-point channels \cite{Ribeiro2012}, interference channels \cite{Ribeiro2012, WeiYu2006, Chen2011a, Naderializadeh2014}, wireless networking \cite{Ribeiro2012, Zhang2006, Wu2013}, as well as multiple access \cite{Wang2010, Rubio2016}, random access \cite{YichuanHu2011, Hu2012} and frequency division multiplexing \cite{Wang2011, Wang2012, He2014}. Less obvious application areas where resource allocation tasks can also be formulated as particular cases of  \eqref{eq:Base_Problem} include MIMO systems \cite{Yu2017a, Shi2011}, beamforming \cite{Sidiropoulos2006,Bazerque2007, Bedi2018}, caching \cite{Bedi2018a}, and wireless control \cite{Gatsis2015, Eisen2019b, Gatsis2018}. Although problems in \cite{Ribeiro2012, Zhang2006, Wu2013, WeiYu2006, Chen2011a, Naderializadeh2014, Wang2010, Rubio2016, YichuanHu2011, Hu2012, Wang2011, Wang2012, He2014, Yu2017a, Shi2011, Sidiropoulos2006, Bazerque2007, Bedi2018, Bedi2018a, Gatsis2015, Eisen2019b, Gatsis2018} have their own difficulties, they all share three challenges that are well-described by \eqref{eq:Base_Problem}: \textit{Dimensionality}, \textit{lack of convexity}, and \textit{model availbaility}. Indeed, when $\ccalH$ is an infinite set --as in most applications-- finding an optimal or near-optimal solution to \eqref{eq:Base_Problem} requires direct policy search, which is a rather obscure and complicated task. Further, while the utilities $g^{o}$ and $\bbg$ and the feasible set $\ccalX$ are often known design choices and can be made concave or convex as needed, this is not the case with the distribution $\ccalM_{\bbH}$, the service metric $\bbf$, or the set $\ccalP$. These entities depend on propagation physics, as well as models of interference and multiple access management. Most often, such models are either inaccurate or unavailable, especially in complex networking settings, whereas in most existing models the form of $\bbf$ and $\ccalP$ render \eqref{eq:Base_Problem} nonconvex \cite{Ribeiro2012}.

Lack of convexity is an inherent challenge and it is accepted that we settle for locally optimal solutions, heuristics, or relaxations. To some extent, the same counts for dimensionality and model availability. However, the recent advent of machine learning for wireless communications \cite{VanChien2019, Xu2019, Lei2017, Sun2018, Lee2018, DeKerret2018, Liang2018, Cui2019, Xu2017, Meng2019, Eisen2019, Eisen2019a} has dawned the realization that both these challenges can be ameliorated with the incorporation of learning parametrizations \cite{Eisen2019, Eisen2019a}. To see why this is true, introduce a parameterization $\boldsymbol{\phi}:{\cal H}\times\mathbb{R}^{N_{\boldsymbol{\phi}}}\rightarrow\mathbb{R}^{N_{R}}$,
and restrict resource allocations as $\bbp(\cdot) \equiv \bbphi(\cdot,\bbtheta), \bbtheta\in\mathbb{R}^{N_{\boldsymbol{\phi}}} $. Then, the base problem \eqref{eq:Base_Problem} may be relaxed as
\begin{equation}
\begin{array}{rl}
\underset{\bbx,\bbtheta}{\mathrm{maximize}} & g^{o}(\bbx)\\
\mathrm{subject\,to} & \bbx\le\mathbb{E}\{\bbf(\bbphi(\bbH,\bbtheta),\bbH)\}\\
& \bbg(\bbx)\ge{\bf 0},\,(\bbx,\bbtheta)\in{\cal X}\times\Theta
\end{array},\label{eq:PFA}
\end{equation}
%
where 
 $\Theta\subseteq\big\{\bbtheta\in\bbR^{N_{\bbphi}}\big|\hspace{0.5bp}\bbphi(\cdot,\bbtheta)\in\ccalP \big\}$ is a nonempty and closed parameter space. Through the parametrization $\bbphi$, also known as a policy function approximation (PFA) \cite{Powell2019}, problem \eqref{eq:PFA} serves as a finite dimensional surrogate for the infinite dimensional problem \eqref{eq:Base_Problem} \cite{Eisen2019}. 
 Solving such a surrogate incurs some inevitable loss of optimality. Nevertheless, this issue may be mitigated by exploiting well-known parametric function classes with universal or near-universal approximation properties such as Radial Basis Functions (RBFs) \cite{Park1991}, Reproducing Kernel Hilbert Spaces (RKHSs) \cite{Sriperumbudur2010} and Deep Neural Networks (DNNs) \cite{Hornik1989}. 

While it is clear that \eqref{eq:PFA} replaces infinite dimensional search by finite dimensional optimization, it is not obvious how \eqref{eq:PFA} can circumvent the need for accurate models. This is addressed in \cite{Eisen2019}, which builds on the observation that the PFA formulation \eqref{eq:PFA} represents a scalarization of a multi-objective statistical learning problem. In fact, each entry of $\bbx$ is associated with an expected reward, with the difference of the two formulating a stochastic constraint. Each expected reward has the form of the objective of a greedy reinforcement learning problem \cite{Powell2011, Bertsekas_Vol_2, Sutton2018,Powell2019}, in which $\bbH$ and $\bbphi(\bbH,\bbtheta)$ correspond to the state and control actions, respectively. In that sense, it is not only that we can reformulate optimal allocation of resources in wireless systems as a learning problem, but that learning resource allocations is \textit{inherently} a learning problem. This observation led to a primal-dual training method for finding an optimal solution to \eqref{eq:PFA} in \cite{Eisen2019}, which relies on stochastic approximation \cite{Kushner2003,Spall2003a}, 
and attains model-free operation borrowing randomization ideas from policy gradient methods in reinforcement learning \cite{Sutton2018}. 

Although the primal-dual learning algorithm of \cite{Eisen2019} has been shown to work well in some examples, including large scale networks with proper parameterizations \cite{Eisen2019a}, issues associated with model-free operation are not addressed. As is the case with policy gradient, the algorithm of \cite{Eisen2019} requires use of randomized policies. We know that these are inefficient as compared with deterministic policies, but we lack understanding of the loss of optimality associated with specific randomization choices. The main contribution of this paper is to put forth a \textit{principled approach} for solving the PFA \eqref{eq:PFA} via model-free training. We do so by avoiding the use of randomized policies altogether, and instead relying on appropriately constructed, smoothed surrogates to \eqref{eq:PFA}, which enable exact zeroth-order gradient representation  \cite{Nesterov2017}. This approach not only yields a new, efficient \textit{and} technically grounded model-free training algorithm, but also enables detailed analysis, quantifying the relation of \textit{both} problems \eqref{eq:Base_Problem} \textit{and}  \eqref{eq:PFA} to the smoothed surrogate corresponding to the latter,
in both primal and dual domains.
Specifically, our contributions are as follows.


\vspace{2pt}\noindent\textbf{The Primal Smoothed Surrogate (Section \ref{sec:Smoothed-Constrained-Program}).} We introduce a new \textit{smoothed surrogate} to the \textit{constrained} parameterized problem \eqref{eq:PFA}, for which we establish consistency, as well as explicit approximation rates. 
Our construction leverages recent results on function approximation via
Gaussian convolution \cite{Nesterov2017}, 
and ensures that both the objective
\textit{and} constraints of the proposed smoothed surrogate approximate those of \eqref{eq:PFA} \textit{uniformly}
in their feasible sets, under mild regularity conditions (Lemmata \ref{lem:gm_hm} and \ref{lem:gm_hm-1}). The quality
of the approximation is controlled by user-prescribed, nonnegative
smoothing parameters $\mu_{S}$ and $\mu_{R}$, each associated with
the decision variables $\bbx$ and $\bbtheta$
of \eqref{eq:PFA}, respectively. 
The proposed surrogate exhibits rather desirable properties.
%
First, as either of the smoothing parameters decreases, the corresponding approximation errors shrink, and \textit{at a linear rate}. 
Second, all smoothed approximations involved are always differentiable, and their gradients may be represented \textit{exactly}
as averages of finite differences, which are uniformly stable relative
to both $\mu_{S}$ and $\mu_{R}$. 
Consequently, such approximations
can be exploited to define \textit{zeroth-order stochastic quasi-gradients} of the objective and all constraints of \eqref{eq:PFA}, with consistent
and predictable behavior. 
Third, it is possible to establish simple and easily satisfiable
conditions \textit{on} \eqref{eq:PFA}, which ensure well-definiteness
and consistency of the smoothed surrogate, as well as feasibility \textit{within}
the feasible sets of both \eqref{eq:PFA} and \eqref{eq:Base_Problem} (Theorems \ref{thm:S_Feasibility} and \ref{thm:S_Feasibility-2}).


\vspace{2pt}\noindent\textbf{The Dual Smoothed Surrogate (Section \ref{subsec:Dual-Optimal-Values}).} We analyze the dual of our smoothed surrogate
\textit{as a smoothed approximation to the dual of} \eqref{eq:PFA}.
We establish explicit upper and lower bounds on the
difference of the respective dual optimal values, with both bounds
being \textit{linearly decreasing relative to both} $\mu_{S}$
\textit{and} $\mu_{R}$ (Theorem \ref{thm:Dual_Approximation}). This result is of independent interest, because
it is the first to confirm that Gaussian smoothing can be effectively
leveraged in the dual domain the design of general zeroth-order (model-free)
methods, applicable to \textit{constrained} programs and, more broadly,
problems of the saddle point type.


\vspace{2pt}\noindent\textbf{Duality Gap of Smoothed Surrogates (Section \ref{subsec:Strong-Duality}).} Assuming an $\epsilon$-universal policy parameterization,
we take \cite{Eisen2019} strictly one step further by completely
characterizing the duality gap between the optimal value of the variational
problem \eqref{eq:Base_Problem} and the dual value of the proposed
smoothed surrogate. Specifically, we show that the aforementioned
duality gap is, in absolute value, of the order of ${\cal O}(\mu_{S}{\textstyle \sqrt{N_{S}}}\hspace{-0.5pt}+\hspace{-0.5pt}\mu_{R}{\textstyle \sqrt{N_{\bbphi}}}\hspace{-0.5pt}+\hspace{-0.5pt}\epsilon)$ (Theorem \ref{thm:Dual_Gap_Smoothed}). 
If $\mu_S\equiv \mu_R\equiv 0$,
our duality result recovers exactly that developed earlier in  \cite{Eisen2019}, whereas, for $\mu_S > 0$ and $\mu_R > 0$, it  explicitly quantifies the combined effects of \textit{both}
policy parameterization \textit{and} smoothing on approximating the
optimal value of the original problem \eqref{eq:Base_Problem} via
surrogate dualization.


\vspace{2pt}\noindent\textbf{Model-Free Learning (Section \ref{sec:Primal-Dual-Model-Free-Learning}).} We develop a new randomized zeroth-order primal-dual
algorithm for tackling \eqref{eq:PFA}, which exploits the stochastic zeroth-order gradient representation of our proposed smoothed approximations, and fits the desired model-free setting \textit{by construction}. 
Our primal-dual algorithm is similar to that proposed in \cite{Eisen2019},
\textit{but with a couple of twists}; it takes advantage of our sensitivity
and duality analyses and, compared to the policy gradient approach
of \cite{Eisen2019}, it requires no policy randomization, and it
operates exclusively on probing $g^{o}$, $\bbg$ and the
\textit{composition} $\bbf(\bbphi(\bbH,\cdot),\bbH)$,
\textit{without} the need of computing the gradient of the parametric
representation $\bbphi(\bbH,\cdot)$. Further,
the proposed algorithm converges at least to a stationary point of
the dualized smoothed surrogate, which satisfies our duality gap guarantees;
its optimal value can be made \textit{arbitrarily close} to the optimal
value of the original resource allocation problem \eqref{eq:Base_Problem}
\textit{at will}, by properly selecting smoothing parameters $\mu_{S}$
and $\mu_{R}$, as well as an $\epsilon$-universal parameterization
$\bbphi$.

\vspace{2pt}\noindent Our contributions are also supported by indicative numerical simulations (Section \ref{sec:Numerical-Simulations-=000026}),
justifying our approach and confirming our theoretical
findings. Indeed, our simulations demonstrate near-ideal performance
of the proposed model-free method, as compared to both strictly optimal
solutions and state-of-the-art heuristics, both relying on availability
of explicit system models.

In the analysis that follows, we assume that the feasible set of \eqref{eq:Base_Problem}
is nonempty, that $\mbE\{\bbf(\bbp(\bbH),\bbH)\}$
exists and is finite for every $\bbp\in\ccalP $, and that
 the optimal value of \eqref{eq:Base_Problem}, $\mathscr{P}^{*}\in(-\infty,\infty]$,
is attained for at least one feasible decision; thus, $\mathscr{P}^{*}<\infty$.
Similar to \eqref{eq:Base_Problem}, we assume that \eqref{eq:PFA}
has at least one feasible point, as well. Then, if $\mathscr{P}_{\bbphi}^{*}\in(-\infty,\infty]$
denotes the optimal value of \eqref{eq:PFA}, it follows that $\mathscr{P}_{\bbphi}^{*}\le\mathscr{P}^{*}$,
implying that $\mathscr{P}_{\bbphi}^{*}<\infty$. For
simplicity, we also assume that $\mathscr{P}_{\bbphi}^{*}$
is attained within the feasible set of \eqref{eq:PFA}.


%
\section{\label{sec:Smoothing-via-Gaussian}Smoothing via Gaussian Convolution}

This section introduces Gaussian smoothing and its properties, and
follows closely the corresponding treatment in \cite{Nesterov2017}.

Let $f:\bbR^{N}\to\bbR$ be Borel. Also, for any
random element $\boldsymbol{U}\hspace{-1pt}:\hspace{-1pt}\Omega\hspace{-1pt}\to\hspace{-1pt}\bbR^{N}$
following the standard Gaussian measure on $\bbR^{N}$, hereafter
denoted as $\boldsymbol{U}\sim{\cal N}\left({\bf 0},\boldsymbol{I}_{N}\right)$,
and for $\mu\ge0$, consider another Borel function $f_{\mu}:\bbR^{N}\to\bbR$,
defined, for every $\bbx\in\bbR^{N}$, as
\begin{equation}
f_{\mu}\left(\bbx\right)\triangleq\mbE\left\{ f\left(\bbx+\mu\boldsymbol{U}\right)\right\} \equiv\int f\left(\bbx+\mu\boldsymbol{u}\right){\cal N}\hspace{-1pt}\left(\boldsymbol{u}\right)\text{d}\boldsymbol{u},
\end{equation}
with ${\cal N}:\bbR^{N}\to\bbR$ being the standard
Gaussian density, i.e.,
\begin{equation}
{\cal N}\hspace{-1pt}\left(\boldsymbol{u}\right)\triangleq\left(2\pi\right)^{-N/2}\exp\bigl(-\left\Vert \boldsymbol{u}\right\Vert _{2}^{2}/2\bigr),\quad\boldsymbol{u}\in\bbR^{N},
\end{equation}
provided that the involved integral is well-defined. For every $\mu>0$,
$f_{\mu}$ may be easily shown to be a convolution of the original
function $f$ with the Gaussian density on $\bbR^{N}$ with
mean zero and covariance equal to $\mu^{2}\boldsymbol{I}_{N}$. Indeed,
for every $\bbx\in\bbR^{N}$, and via a simple change
of variables, it is true that
\begin{flalign}
f_{\mu}\left(\bbx\right) & =\int f\left(\boldsymbol{u}\right)\mu^{-N}{\cal N}\hspace{-1pt}\left(\dfrac{\bbx-\boldsymbol{u}}{\mu}\right)\text{d}\boldsymbol{u}\nonumber \\
 & \equiv\bigl(f*\big[\mu^{-N}{\cal N}\bigl(\left(\cdot\right)\mu^{-1}\bigr)\big]\bigr)\left(\bbx\right).
\end{flalign}
Therefore, the smoothed function $f_{\mu}$ may be seen as the output
of a linear filter whose impulse response is the standard gaussian
pulse, taking $f$ as its input.

In many cases, $f_{\mu}$ turns out to be everywhere differentiable
on $\bbR^{N}$, \textit{even if $f$ is not}, whereas the gradient
of $f_{\mu}$ admits a \textit{zeroth-order representation}. In particular,
such is the case of all Lipschitz functions on $\bbR^{N}$ \cite{Nesterov2017},
as the next result suggests. 
\begin{lem}
\textbf{\textup{(Properties of $f_{\mu}$ \cite{Nesterov2017})}}\label{lem:Grad_FM}
Let $\boldsymbol{U}\sim{\cal N}\left({\bf 0},\boldsymbol{I}_{N}\right)$,
and consider any globally Lipschitz function $f:\bbR^{N}\to\bbR$.
Then, for any ${\cal F}\subseteq\bbR^{N}$, the following statements
are true:
\begin{itemize}
\item For every $\mu\ge0$, $f_{\mu}$ is well-defined and finite on ${\cal F}$,
and
\begin{equation}
\sup_{\bbx\in{\cal F}}|f_{\mu}\left(\bbx\right)-f\left(\bbx\right)\hspace{-1pt}\hspace{-1pt}|\le\mu L\sqrt{N}.\label{eq:Value_Approximation}
\end{equation}
\end{itemize}
\vspace{-4bp}

\begin{itemize}
\item If $f$ is convex on $\bbR^{N}$, so is $f_{\mu}$, and $f_{\mu}\ge f$
on ${\cal F}$.
\item For every $\mu>0$, $f_{\mu}$ is differentiable on ${\cal F}$, and
its gradient $\nabla f_{\mu}:\bbR^{N}\to\bbR^{N}$
admits the representation
\begin{align}
\hspace{-1pt}\hspace{-1pt}\hspace{-1pt}\nabla f_{\mu}\left(\bbx\right) & \equiv\mbE\bigg\{\hspace{-0.5pt}\hspace{-0.5pt}\dfrac{f\left(\bbx+\mu\boldsymbol{U}\right)-f\left(\bbx\right)}{\mu}\boldsymbol{U}\hspace{-0.5pt}\hspace{-0.5pt}\bigg\},\label{eq:HELP_1}
\end{align}
for all $\bbx\in{\cal F}$. Further, it is true that
\begin{align}
\hspace{-0.5pt}\hspace{-0.5pt}\hspace{-0.5pt}\hspace{-0.5pt}\hspace{-0.5pt}\hspace{-0.5pt}\hspace{-0.5pt}\hspace{-0.5pt}\hspace{-0.5pt}\hspace{-0.5pt}\hspace{-0.5pt}\hspace{-0.5pt}\sup_{\bbx\in{\cal F}}\mbE\hspace{-0.5pt}\hspace{-0.5pt}\left\{ \bigg\Vert\dfrac{f\left(\bbx\hspace{-0.5pt}+\hspace{-0.5pt}\mu\boldsymbol{U}\right)\hspace{-0.5pt}-\hspace{-0.5pt}f\left(\bbx\right)}{\mu}\boldsymbol{U}\bigg\Vert_{2}^{2}\right\}  & \hspace{-0.5pt}\hspace{-0.5pt}\le\hspace{-0.5pt}\hspace{-0.5pt}L^{2}(N\hspace{-0.5pt}\hspace{-0.5pt}+\hspace{-0.5pt}4)^{2}.\hspace{-0.5pt}\hspace{-0.5pt}\hspace{-0.5pt}\hspace{-0.5pt}\hspace{-0.5pt}\hspace{-0.5pt}\hspace{-0.5pt}\hspace{-0.5pt}\label{eq:Grad_Approximation}
\end{align}
\end{itemize}
\end{lem}
Lemma \ref{lem:Grad_FM} will be key to the results presented in this
paper, as discussed in detail as follows.

\section{\label{sec:Smoothed-Constrained-Program}Smoothed Constrained Program
Surrogates}

In this section, we introduce a new, \textit{smoothed surrogate} of
the \textit{whole} constrained program \eqref{eq:PFA}, as promised
in Section \ref{sec:Optimal-Resource-Allocation}, leveraging
the results of Section \ref{sec:Smoothing-via-Gaussian}. We also
introduce conditions under which this smoothed surrogate is well-defined,
and establish various of its properties, as well as its structural
relation to \eqref{eq:PFA}. The power of the proposed surrogate is
in that it provides a technically grounded means for dealing with
\eqref{eq:PFA} in the model-free setting, i.e., when the functions
$g^{o}$, $\bbg$ and $\bbf$ are apriori unknown,
and may be only observed through \textit{limited} \textit{probing}.

\subsection{Surrogate Construction}

Let $\mu_{S}\ge0$, $\mu_{R}\ge0$, and consider random elements $\boldsymbol{U}_{S}\hspace{-0.5pt}\sim\hspace{-0.5pt}\hspace{-0.5pt}{\cal N}(\boldsymbol{0},\boldsymbol{I}_{N_{S}})$
and $\boldsymbol{U}_{R}\hspace{-0.5pt}\hspace{-0.5pt}\hspace{-0.5pt}\sim\hspace{-0.5pt}\hspace{-0.5pt}{\cal N}(\boldsymbol{0},\boldsymbol{I}_{N_{\bbphi}})$,
the latter taken independent of $\bbH$. Driven by the results
of Section \ref{sec:Smoothing-via-Gaussian}, we define smoothed versions
of $g^{o}$, $\bbg$ and $\mbE\{\bbf(\bbphi(\bbH,\cdot),\hspace{-0.5pt}\bbH)\}\hspace{-0.5pt}\hspace{-0.5pt}\hspace{-0.5pt}\triangleq\hspace{-0.5pt}\hspace{-0.5pt}\hspace{-0.5pt}\overline{\bbf}^{\bbphi}(\cdot)$
as
\begin{flalign}
g_{\mu_{S}}^{o}(\bbx) & \triangleq\mbE\{g^{o}(\bbx+\mu_{S}\boldsymbol{U}_{S})\},\quad\bbx\in\ccalX,\\
\bbg_{\mu_{S}}(\bbx) & \triangleq\mbE\{\bbg(\bbx+\mu_{S}\boldsymbol{U}_{S})\},\quad\bbx\in\ccalX\quad\text{and}\\
\overline{\bbf}_{\mu_{R}}^{\bbphi}(\bbtheta) & \triangleq\mbE\{\bbf(\bbphi(\bbH,\bbtheta+\mu_{R}\boldsymbol{U}_{R}),\bbH)\},\quad\bbtheta\in\Theta,
\end{flalign}
where, at this point, we arbitrarily assume that the involved expectations
are well-defined and finite on $\ccalX$ and $\Theta$. We will
return to those issues shortly. Then, we may formulate a (hopefully)
smoothed version of problem \eqref{eq:PFA} as\vspace{-2bp}
\begin{equation}
\begin{array}{rl}
\underset{\bbx,\bbp(\cdot)}{\mathrm{maximize}}  & g_{\mu_{S}}^{o}(\bbx)\\
\mathrm{subject\,to} & \bbx+\mathtt{S}(\mu_{R})\le\overline{\bbf}_{\mu_{R}}^{\bbphi}(\bbtheta)\\
 & \bbg_{\mu_{S}}(\bbx)\ge{\bf 0},\,(\bbx,\bbtheta)\in\ccalX\times\Theta
\end{array},\label{eq:Zero_PFA}
\end{equation}
where $\mathtt{S}:\bbR_{+}\to\bbR_{+}^{N_{S}}$
is a nonnegative \textit{feasibility slack}, with properties to be
determined. Formulation of the smoothed surrogate \eqref{eq:Zero_PFA}
is well motivated due to the fact that, whenever the objective $g^{o}$
and all entries of the constraint vector functions $\bbg$
and $\bbf(\bbphi(\bbH,\cdot),\hspace{-0.5pt}\bbH)$
are sufficiently well-behaved, such that Lemma \ref{lem:Grad_FM}
appropriately applies, the smoothed functions $g_{\mu_{S}}^{o}$,
$\bbg_{\mu_{S}}$ and $\overline{\bbf}_{\mu_{R}}^{\bbphi}$
are differentiable, and the respective gradients may be represented
as averages of suitably defined finite differences. This is particularly
important in developing effective and predictable methods for solving
problem \eqref{eq:PFA} in the model-free setting: Finite differences
are by construction based on function evaluations only. Thus,
the surrogate \eqref{eq:Zero_PFA} constitutes a natural \textit{zeroth-order
proxy} for dealing with the original parameterized problem \eqref{eq:PFA}.

However, before focusing on how to use \eqref{eq:Zero_PFA} in order
to solve \eqref{eq:PFA}, we have to make sure that \eqref{eq:Zero_PFA}
is a well-defined \textit{and} feasible problem, and also reveal its
fundamental connection to \eqref{eq:PFA}. These tasks are the subject
of the rest of this section.

\subsection{Smoothing $g^{o},\bbg$ and $\overline{\bbf}^{\bbphi}$}

Our treatment will require imposing appropriate structure on the functions
involved in \eqref{eq:PFA}, as we now discuss in detail. Hereafter,
the $i$-th entries of $\bbg$ (resp. $\bbg_{\mu_{S}}$)
and $\bbf$ (resp. $\overline{\bbf}_{\mu_{R}}^{\bbphi}$)
will be denoted as $g^{i}$ ($g_{\mu_{S}}^{i}$), $i\hspace{-0.5pt}\hspace{-0.5pt}\hspace{-0.5pt}\in\hspace{-0.5pt}\hspace{-0.5pt}\hspace{-0.5pt}\mathbb{N}_{N_{\bbg}}^{+}$
and $f^{i}$ ($\overline{f}_{\mu_{R}}^{\bbphi,i}$), $i\hspace{-0.5pt}\hspace{-0.5pt}\hspace{-0.5pt}\in\hspace{-0.5pt}\hspace{-0.5pt}\hspace{-0.5pt}\mathbb{N}_{N_{S}}^{+}$,
respectively.\setlist[description]{style=multiline}

\noindent \begin{assumption}\label{AssumptionMain}The following
conditions are satisfied:
\begin{description}
\item [{\textmd{${\bf C1}$}}] \noindent For every $i\hspace{-0.5pt}\hspace{-0.5pt}\hspace{-0.5pt}\in\hspace{-0.5pt}\hspace{-0.5pt}\hspace{-0.5pt}\hspace{-0.5pt}\{o,\hspace{-0.5pt}\mathbb{N}_{N_{\bbg}}^{+}\hspace{-0.5pt}\}\hspace{-0.5pt}$,
$g^{i}$ is $L_{g}^{i}$-Lipschitz on $\bbR^{N_{S}}$.
\item [{\textmd{${\bf C2}$}}] \noindent For every $i\hspace{-0.5pt}\hspace{-0.5pt}\hspace{-0.5pt}\in\hspace{-0.5pt}\hspace{-0.5pt}\hspace{-0.5pt}\mathbb{N}_{N_{S}}^{+}$,
there is $L_{f}^{i}<\infty$, such that\vspace{-6bp}
\begin{multline}
\hspace{6pt}\hspace{-0.5pt}\hspace{-0.5pt}\hspace{-0.5pt}\hspace{-0.5pt}\hspace{-0.5pt}\hspace{-0.5pt}\hspace{-0.5pt}\hspace{-6pt}\Vert f^{i}(\bbphi(\bbH,\bbtheta_{1}),\bbH)-f^{i}(\bbphi(\bbH,\bbtheta_{2}),\bbH)\Vert_{{\cal L}_{2}}\\
\;\le L_{f}^{i}\Vert\bbtheta_{1}-\bbtheta_{2}\Vert_{2},\,\forall\, (\bbtheta_{1},\bbtheta_{2})\in\bbR^{N_{\bbphi}}\times\bbR^{N_{\bbphi}}.\hspace{-0.5pt}\hspace{-0.5pt}\hspace{-0.5pt}\hspace{-0.5pt}\hspace{-0.5pt}\hspace{-0.5pt}\hspace{-0.5pt}\hspace{8pt}
\end{multline}
\end{description}
\end{assumption}
Condition $\mathbf{C2}$ of Assumption \ref{AssumptionMain} has the
following consequences on the behavior of $\mbE\{\bbf(\bbphi(\bbH,\cdot),\bbH)\}\equiv\overline{\bbf}^{\bbphi}(\cdot)$.
\begin{prop}
\textbf{\textup{(Properties of $\overline{\bbf}^{\bbphi}$)}}\label{prop:Lipschitz-Properties}
Suppose that condition $\mathbf{C2}$ of Assumption 2 is in effect.
Then, for every $i\in\mathbb{N}_{N_{S}}^{+}$, $\overline{f}^{\bbphi,i}$
is $L_{f}^{i}$-Lipschitz on $\bbR^{N_{\bbphi}}$.
Additionally, it is true that\vspace{-3.5bp}
\begin{multline}
\quad\quad\mbE\{|f^{i}(\bbphi(\bbH,\bbtheta+\boldsymbol{u}),\bbH)|\}\\
\le L_{f}^{i}\Vert\boldsymbol{u}\Vert_{2}+\mbE\{|f^{i}(\bbphi(\bbH,\bbtheta),\bbH)|\},\quad\quad
\end{multline}
for all $\left(\bbtheta,\boldsymbol{u}\right)\in\Theta\times\bbR^{N_{\bbphi}}$,
and for all $i\in\mathbb{N}_{N_{S}}^{+}$.
\end{prop}
\begin{IEEEproof}[Proof of Proposition \ref{prop:Lipschitz-Properties}]
The first part of the result follows immediately from condition ${\bf C2}$,
by the nested structure of ${\cal L}_{p}$-spaces, and Jensen. The
second part follows via an application of the triangle inequality.
\end{IEEEproof}
Assumption \ref{AssumptionMain} and Proposition \ref{prop:Lipschitz-Properties}
may be further exploited to establish well-definiteness and basic
properties of $g_{\mu_{S}}^{o}$, $\bbg_{\mu_{S}}$ and
$\overline{\bbf}_{\mu_{R}}^{\bbphi}$. To this
end, for $\bbx\in\ccalX,\mu_{S}>0$ and for every $i\in\{o,\hspace{-0.5pt}\mathbb{N}_{N_{\bbg}}^{+}\hspace{-0.5pt}\}$,
let us define finite differences\vspace{-2.5bp}
\begin{equation}
\Delta_{g}^{i}(\bbx,\mu_{S},\boldsymbol{U}_{S})\triangleq\dfrac{g^{i}(\bbx+\mu_{S}\boldsymbol{U}_{S})-g^{i}(\bbx)}{\mu_{S}}.\label{eq:Finite_Diff_1}
\end{equation}
Similarly, for $\bbtheta\in\Theta,\mu_{R}>0$ and for every
$i\hspace{-0.5pt}\hspace{-0.5pt}\hspace{-0.5pt}\in\hspace{-0.5pt}\hspace{-0.5pt}\hspace{-0.5pt}\mathbb{N}_{N_{S}}^{+}$,
define\vspace{-4bp}
\begin{multline}
\Delta_{f}^{i}(\bbtheta,\mu_{R},\boldsymbol{U}_{R},\bbH)\\
\triangleq\dfrac{f^{i}(\bbphi(\bbH,\bbtheta+\mu_{R}\boldsymbol{U}_{R}),\bbH)-f^{i}(\bbphi(\bbH,\bbtheta),\bbH)}{\mu_{R}}.\label{eq:Finite_Diff_2}
\end{multline}
The relevant results now follow.
\begin{lem}
\textbf{\textup{(Existence \& Properties of $g_{\mu_{S}}^{o}$ and
$\bbg_{\mu_{S}}$)}}\label{lem:gm_hm} Suppose that Assumption
\ref{AssumptionMain} is in effect. Then, for every $i\in\{o,\hspace{-0.5pt}\mathbb{N}_{N_{\bbg}}^{+}\hspace{-0.5pt}\}$
and for every $\mu_{S}>0$, each $g_{\mu_{S}}^{i}$ is a well-defined,
finite, concave and everywhere differentiable underestimator of $g^i$ on $\ccalX$, such that
\begin{align}
\hspace{-0.5pt}\hspace{-0.5pt}\hspace{-0.5pt}\sup_{\bbx\in\ccalX}|g_{\mu_{S}}^{i}\left(\bbx\right)-g^{i}\left(\bbx\right)\hspace{-1pt}\hspace{-1pt}| & \hspace{-1pt}\le\hspace{-1pt}\mu_{S}L_{g}^{i}{\textstyle \sqrt{N_{S}}},\\
\sup_{\bbx\in\ccalX}\mbE\{\Vert\Delta_{g}^{i}(\bbx,\mu_{S},\boldsymbol{U}_{S})\boldsymbol{U}_{S}\Vert_{2}^{2}\} & \hspace{-1pt}\le\hspace{-1pt}(L_{g}^{i})^{2}(N_{S}\hspace{-0.5pt}\hspace{-0.5pt}\hspace{-0.5pt}+\hspace{-0.5pt}\hspace{-0.5pt}\hspace{-0.5pt}4)^{2}\label{eq:Uni_1}\\
\text{and}\quad\mbE\{\Delta_{g}^{i}(\bbx,\mu_{S},\boldsymbol{U}_{S})\boldsymbol{U}_{S}\} & \hspace{-1pt}\equiv\hspace{-1pt}\nabla g_{\mu_{S}}^{i}\hspace{-0.5pt}\left(\bbx\right),\hspace{-0.5pt}\hspace{-0.5pt}\hspace{-0.5pt}\hspace{-0.5pt}\hspace{-0.5pt}\hspace{-0.5pt}\hspace{-0.5pt}\hspace{-0.5pt}\hspace{-0.5pt}\hspace{-0.5pt}\hspace{-0.5pt}\hspace{-0.5pt}\hspace{-0.5pt}\hspace{-0.5pt}\hspace{-0.5pt}\hspace{-0.5pt}
\end{align}
for all $\bbx\in\ccalX$.
\end{lem}
\begin{IEEEproof}[Proof of Lemma \ref{lem:gm_hm}]
Trivial, see Lemma \ref{lem:Grad_FM} (Section \ref{sec:Smoothing-via-Gaussian}).
\end{IEEEproof}
\begin{lem}
\textbf{\textup{(Existence \& Properties of $\overline{\bbf}_{\mu_{R}}^{\bbphi}$)}}\label{lem:gm_hm-1}
Suppose that Assumption \ref{AssumptionMain} is in effect. Then,
for every $i\hspace{-0.5pt}\hspace{-0.5pt}\hspace{-0.5pt}\in\hspace{-0.5pt}\hspace{-0.5pt}\hspace{-0.5pt}\mathbb{N}_{N_{S}}^{+}$
and for every $\mu_{R}>0$, each $f_{\mu_{R}}^{i}$ is well-defined,
finite, differentiable everywhere on $\Theta$, such that
\begin{flalign}
\sup_{\boldsymbol{\bbtheta}\in\Theta}\big|\overline{f}_{\mu_{R}}^{\bbphi,i}\left(\bbtheta\right)\hspace{-0.5pt}-\hspace{-0.5pt}\overline{f}^{\bbphi,i}\left(\boldsymbol{\bbtheta}\right)\hspace{-1pt}\hspace{-1pt}\big| & \hspace{-1pt}\le\hspace{-1pt}\mu_{R}L_{f}^{i}{\textstyle \sqrt{N_{\bbphi}}},\\
\sup_{\bbtheta\in\Theta}\mbE\{\Vert\Delta_{f}^{i}(\bbtheta,\mu_{R},\boldsymbol{U}_{R},\bbH)\boldsymbol{U}_{R}\Vert_{2}^{2}\} & \hspace{-1pt}\le\hspace{-1pt}(L_{f}^{i})^{2}(N_{\bbphi}\hspace{-0.5pt}\hspace{-0.5pt}\hspace{-0.5pt}+\hspace{-0.5pt}\hspace{-0.5pt}4)^{2}\label{eq:Uni_2}\\
\text{and}\quad\mbE\{\Delta_{f}^{i}(\bbtheta,\mu_{R},\boldsymbol{U}_{R},\bbH)\boldsymbol{U}_{R}\} & \hspace{-1pt}\equiv\hspace{-1pt}\nabla\overline{f}_{\mu_{R}}^{\bbphi,i}\left(\bbtheta\right),
\end{flalign}
for all $\bbtheta\in\Theta$.
\end{lem}
\begin{IEEEproof}[Proof of Lemma \ref{lem:gm_hm-1}]
Fix $i\hspace{-0.5pt}\hspace{-0.5pt}\hspace{-0.5pt}\in\hspace{-0.5pt}\hspace{-0.5pt}\hspace{-0.5pt}\mathbb{N}_{N_{S}}^{+}$,
and consider the function $\overline{f}^{\bbphi,i}(\cdot)\equiv\mbE\{f^{i}(\bbphi(\bbH,\cdot),\hspace{-0.5pt}\bbH)\}$,
which, by Proposition \ref{prop:Lipschitz-Properties}, is \textit{$L_{f}^{i}$}-Lipschitz
on $\bbR^{N_{\bbphi}}$. Then, Lemma \ref{lem:Grad_FM}
implies that, for every $\mu_{R}\ge0$,
\begin{equation}
\sup_{\boldsymbol{\bbtheta}\in\Theta}\big|\mbE\big\{\overline{f}^{\bbphi,i}\left(\bbtheta+\mu_{R}\boldsymbol{U}_{R}\right)\hspace{-1pt}\hspace{-1pt}\big\}\hspace{-0.5pt}-\hspace{-0.5pt}\overline{f}^{\bbphi,i}\left(\boldsymbol{\bbtheta}\right)\hspace{-1pt}\hspace{-1pt}\big|\le\mu_{R}L_{f}^{i}{\textstyle \sqrt{N_{\bbphi}}}.
\end{equation}
Note that we are not done yet, since $\mbE\big\{\overline{f}^{\bbphi,i}\left((\cdot)+\mu_{R}\boldsymbol{U}_{R}\right)\hspace{-0.5pt}\hspace{-0.5pt}\hspace{-0.5pt}\hspace{-0.5pt}\big\}$
involves an iterated expectation, and \textit{not} an expectation
relative to the joint distribution of $\boldsymbol{U}_{R}$ and $\bbH$.
However, again by Proposition \ref{prop:Lipschitz-Properties}, it
follows that, for every $\bbtheta\in\Theta$,\vspace{-2.5bp}
\begin{flalign}
 & \hspace{-1pt}\hspace{-1pt}\hspace{-1pt}\hspace{-1pt}\hspace{-1pt}\hspace{-1pt}\hspace{-1pt}\hspace{-1pt}\hspace{-1pt}\int\mbE\{|f^{i}(\bbphi(\bbH,\bbtheta+\mu_{R}\boldsymbol{u}),\bbH)|\}\ccalP _{\boldsymbol{U}_{R}}\hspace{-1pt}\left(\mathrm{d}\boldsymbol{u}\right)\nonumber \\
 & \le\mu_{R}L_{f}^{i}\int\Vert\boldsymbol{u}\Vert_{2}\ccalP _{\boldsymbol{U}_{R}}\hspace{-1pt}\left(\mathrm{d}\boldsymbol{u}\right)+\mbE\{|f^{i}(\bbphi(\bbH,\bbtheta),\bbH)|\}\nonumber \\
 & \le\mu_{R}L_{f}^{i}\sqrt{N_{\bbphi}}+\mbE\{|f^{i}(\bbphi(\bbH,\bbtheta),\bbH)|\}<\infty.
\end{flalign}
Then, Fubini's Theorem (Corollary 2.6.5 and Theorem 2.6.6 in \cite{Ash2000Probability})
implies that $\overline{f}_{\mu_{R}}^{\bbphi,i}\left(\cdot\right)\equiv\mbE\{f^{i}(\bbphi(\bbH,(\cdot)+\mu_{R}\boldsymbol{U}_{R}),\bbH)\}$
is finite on $\Theta$, and that\vspace{-2bp}
\begin{flalign}
 & \hspace{-1pt}\hspace{-1pt}\hspace{-1pt}\hspace{-1pt}\hspace{-1pt}\hspace{-1pt}\mbE\big\{\overline{f}^{\bbphi,i}\left(\bbtheta+\mu_{R}\boldsymbol{U}_{R}\right)\hspace{-1pt}\hspace{-1pt}\big\}\nonumber \\
 & \equiv\int\mbE\{f^{i}(\bbphi(\bbH,\bbtheta+\mu_{R}\boldsymbol{u}),\bbH)\}\ccalP _{\boldsymbol{U}_{R}}\hspace{-1pt}\left(\mathrm{d}\boldsymbol{u}\right)\nonumber \\
 & \equiv\int f^{i}(\bbphi(\bbH,\bbtheta+\mu_{R}\boldsymbol{u}),\bbH)\left[\ccalP _{\bbH}\times\ccalP _{\boldsymbol{U}_{R}}\right]\left(\mathrm{d}\left[\boldsymbol{u},\bbH\right]\right)\nonumber \\
 & \equiv\overline{f}_{\mu_{R}}^{\bbphi,i}\left(\bbtheta\right),\quad\forall\,\bbtheta\in\Theta,\label{eq:Fubini_Wonder}
\end{flalign}
where $\bbH$ and $\boldsymbol{U}_{R}$ are statistically
independent by assumption; now we are done. Next, differentiability
of $\overline{f}_{\mu_{R}}^{\bbphi,i}$, as well as the
form of its gradient also follow from Lemma \ref{lem:Grad_FM} on
$\overline{f}^{\bbphi,i}$ and, again, \eqref{eq:Fubini_Wonder}.
Finally, to verify \eqref{eq:Uni_2}, we may write (due to condition
${\bf C2}$)\vspace{-2bp}
\begin{flalign}
 & \hspace{-1pt}\hspace{-1pt}\hspace{-1pt}\hspace{-1pt}\mbE\{\Vert\Delta_{f}^{i}(\bbtheta,\mu_{R},\boldsymbol{U}_{R},\bbH)\boldsymbol{U}_{R}\Vert_{2}^{2}\}\nonumber \\
 & \equiv\dfrac{1}{\mu_{R}^{2}}\mbE\big\{\mbE\big\{|f^{i}(\bbphi(\bbH,\bbtheta+\mu_{R}\boldsymbol{U}_{R}),\bbH)\nonumber \\
 & \quad\quad\quad-\hspace{-1pt}\hspace{-1pt}f^{i}(\bbphi(\bbH,\bbtheta),\bbH)|^{2}|\boldsymbol{U}_{R}\big\}\hspace{-1pt}\hspace{-1pt}\left\Vert \boldsymbol{U}_{R}\right\Vert _{2}^{2}\hspace{-1pt}\big\}\nonumber \\
 & \le\dfrac{1}{\mu_{R}^{2}}\mbE\big\{\hspace{-1pt}(L_{f}^{i})^{2}\Vert\mu_{R}\boldsymbol{U}_{R}\Vert_{2}^{2}\hspace{-1pt}\hspace{-1pt}\left\Vert \boldsymbol{U}_{R}\right\Vert _{2}^{2}\hspace{-1pt}\big\}\nonumber \\
 & \le(L_{f}^{i})^{2}(N_{\bbphi}\hspace{-0.5pt}\hspace{-0.5pt}\hspace{-0.5pt}+\hspace{-0.5pt}\hspace{-0.5pt}4)^{2},
\end{flalign}
as required. The proof is complete.
\end{IEEEproof}
\begin{rem}
We would like to mention that a weaker version of Lemma \ref{lem:gm_hm}
holds if we weaken condition ${\bf C2}$ of Assumption \ref{AssumptionMain},
by replacing the ${\cal L}_{2}$-norm with an ${\cal L}_{1}$-norm.
In this case, the $\ell_{2}$-norm-squared inside the expectation
of \eqref{eq:Uni_2} would be replaced by a mere $\ell_{2}$-norm;
essentially, only boundedness of $\Vert\Delta_{f}^{i}(\bbtheta,\mu_{R},\boldsymbol{U}_{R},\bbH)\boldsymbol{U}_{R}\Vert_{2}$
in ${\cal L}_{1}$ would be guaranteed, instead of boundedness in
${\cal L}_{2}$. The main reason why \eqref{eq:Uni_2} (and therefore
condition ${\bf C2}$) is desirable is that it crucially affects the
behavior of gradient-based algorithms for solving problems such as
\eqref{eq:Zero_PFA}, considered later in this work.\hfill{}\Halmos
\end{rem}

\subsection{Surrogate Feasibility}

We are now in place to investigate conditions ensuring feasibility
of the smoothed surrogate \eqref{eq:Zero_PFA}. In particular, we
will be interested in conditions ensuring feasibility of \eqref{eq:Zero_PFA},
\textit{but are on the original parameterized problem \eqref{eq:PFA}}.
This is very important from a practical point of view, since the exact
form of \eqref{eq:Zero_PFA} will be, in most cases, unknown. On the
other hand, feasibility of \eqref{eq:PFA} should be somehow guaranteed
apriori, \textit{even in the model-free setting}; indeed, \textit{both}
resource allocation problems \eqref{eq:Base_Problem} and \eqref{eq:PFA}
are initially proposed by the wireless engineer, who is the one responsible
for formulating meaningful resource allocation tasks.

It turns out that all that is needed for \eqref{eq:Zero_PFA} to be
feasible is the existence of at least one \textit{strictly} feasible
point for \eqref{eq:PFA}. What is more, \eqref{eq:Zero_PFA} can
be made strictly feasible at will. The relevant result follows right
after we define the vectors
\begin{equation}
\boldsymbol{c}_{S}\triangleq\big[L_{g}^{1}\,\ldots\,L_{g}^{N_{\bbg}}\big]^{\hspace{-0.5pt}\boldsymbol{T}}\quad\text{and}\quad\boldsymbol{c}_{R}\triangleq\big[L_{f}^{1}\,\ldots\,L_{f}^{N_{S}}\big]^{\hspace{-0.5pt}\boldsymbol{T}},
\end{equation}
and under the following assumption.

\noindent \begin{assumption}\label{AssumptionSlack}The feasibility
slack $\mathtt{S}_{\bbf}$ is increasing around the origin,
and $\lim_{\mu_{R}\downarrow0}\mathtt{S}_{\bbf}(\mu_{R})\equiv\mathtt{S}_{\bbf}(0)\equiv{\bf 0}$.\end{assumption}
\begin{thm}
\textbf{\textup{(Surrogate Strict Feasibility)}}\label{thm:S_Feasibility}
Let Assumptions \ref{AssumptionMain} and \ref{AssumptionSlack} be
in effect, and suppose that $(\bbx^{\dagger},\bbtheta^{\dagger})\in\bbR^{N_{S}}\times\bbR^{N_{\bbphi}}$
is a strictly feasible point of the parameterized problem \eqref{eq:PFA}.
Then there exist $\mu_{S}^{\dagger}>0$ and $\mu_{R}^{\dagger}>0$,
possibly dependent on $(\bbx^{\dagger},\bbtheta^{\dagger})$,
such that, for every $0\le\mu_{S}\le\mu_{S}^{\dagger}$ and $0\le\mu_{R}\le\mu_{R}^{\dagger}$,
the same point $(\bbx^{\dagger},\bbtheta^{\dagger})$
is strictly feasible for the smoothed surrogate \eqref{eq:Zero_PFA}.
\end{thm}
\begin{IEEEproof}[Proof of Theorem \ref{thm:S_Feasibility}]
Let the point $(\bbx^{\dagger},\bbtheta^{\dagger})\in\bbR^{N_{S}}\times\bbR^{N_{\bbphi}}$
be strictly feasible for \eqref{eq:PFA}, implying that $(\bbx^{\dagger},\bbtheta^{\dagger})\in\ccalX\times\Theta$,
and 
\begin{equation}
\bbg(\bbx^{\dagger})\ge\mathtt{s}_{\bbg}^{\dagger}\quad\text{and}\quad\overline{\bbf}^{\bbphi}(\bbtheta^{\dagger})-\bbx^{\dagger}\ge\mathtt{s}_{\bbf}^{\dagger},
\end{equation}
for some positive slacks $\mathtt{s}_{\bbg}^{\dagger}\in\bbR_{++}^{N_{\bbg}}$
and $\mathtt{s}_{\bbf}^{\dagger}\in\bbR_{++}^{N_{S}}$.
Also, from Lemma \ref{lem:gm_hm}, it follows that, for every $\mu_{S}\ge0$
and $\mu_{R}\ge0$,
\begin{flalign}
\bbg(\bbx^{\dagger}) & \le\bbg_{\mu_{S}}(\bbx^{\dagger})+\mu_{S}\boldsymbol{c}_{S}{\textstyle \sqrt{N_{S}}}\quad\text{and}\\
\overline{\bbf}^{\bbphi}(\bbtheta^{\dagger}) & \le\overline{\bbf}_{\mu_{R}}^{\bbphi}(\bbtheta^{\dagger})+\mu_{R}\boldsymbol{c}_{R}{\textstyle \sqrt{N_{\bbphi}}},
\end{flalign}
Consequently, it is true that
\begin{flalign}
\bbg_{\mu_{S}}(\bbx^{\dagger}) & \ge\mathtt{s}_{\bbg}^{\dagger}-\mu_{S}\boldsymbol{c}_{S}{\textstyle \sqrt{N_{S}}}\quad\text{and}\\
\overline{\bbf}_{\mu_{R}}^{\bbphi}(\bbtheta^{\dagger})-\bbx^{\dagger} & \ge\mathtt{s}_{\bbf}^{\dagger}-\mu_{R}\boldsymbol{c}_{R}{\textstyle \sqrt{N_{\bbphi}}}.
\end{flalign}
Therefore, we can find $\mu_{S}^{\dagger}>0$ and $\mu_{R}^{\dagger}>0$
\textit{sufficiently small but strictly positive}, such that, for
every $0\le\mu_{S}\le\mu_{S}^{\dagger}$ and $0\le\mu_{R}\le\mu_{R}^{\dagger}$,
the strict inequalities $\mathtt{s}_{\bbg}^{\dagger}-\mu_{S}\boldsymbol{c}_{S}\sqrt{N_{S}}>{\bf 0}$
and $\mathtt{s}_{\bbf}^{\dagger}-\mu_{R}\boldsymbol{c}_{R}\sqrt{N_{\bbphi}}>\mathtt{S}(\mu_{R})$
hold. This, of course, implies that $(\bbx^{\dagger},\bbtheta^{\dagger})$
is a strictly feasible point for problem \eqref{eq:Zero_PFA}, for
all aforementioned choices of $\mu_{S}$ and $\mu_{R}$.
\end{IEEEproof}
Theorem \ref{thm:S_Feasibility} is important, as it confirms
the existence of a strictly feasible point for problem \eqref{eq:Zero_PFA},
\textit{uniformly relative to $\mu_{S}$ and }$\mu_{R}$, the latter
being allowed to vary in appropriate sets, whose length is controlled
by the feasibility of \eqref{eq:PFA} and the feasibility slack of
\eqref{eq:Zero_PFA}. An evident byproduct of Theorem \ref{thm:S_Feasibility}
is that \eqref{eq:Zero_PFA} is a feasible and, therefore, meaningful
optimization problem.

Another similar question we may ask is how much the constraints of
\eqref{eq:PFA} are violated for every feasible solution of \eqref{eq:Zero_PFA}.
In this respect, we may formulate the following result.
\begin{thm}
\textbf{\textup{(PFA Constraint Violation)}}\label{thm:S_Feasibility-2}
Let Assumption \ref{AssumptionMain} be in effect. Then, for every
$\mu_{R}\ge0$ such that
\begin{flalign}
\mathtt{S}(\mu_{R})-\mu_{R}\boldsymbol{c}_{R}{\textstyle \sqrt{N_{\bbphi}}} & \ge{\bf 0},\label{eq:FeasCond_2}
\end{flalign}
and for every $\mu_{S}\ge0$, every feasible point of \eqref{eq:Zero_PFA}
is also feasible for \eqref{eq:PFA}. Otherwise, if \eqref{eq:FeasCond_2}
fails to hold, then the negative values of its left-hand-side correspond
to the respective levels of maximal constraint violation for \eqref{eq:PFA}.
\end{thm}
\begin{IEEEproof}[Proof of Theorem \ref{thm:S_Feasibility-2}]
Fix qualifying $\mu_{S}$ and $\mu_{R}$, and let the point $(\bbx^{\mu_{S}},\bbtheta^{\mu_{R}})\in\bbR^{N_{S}}\times\bbR^{N_{\bbphi}}$
be feasible for problem \eqref{eq:Zero_PFA}. Then, it is in fact
true that $(\bbx^{\mu_{S}},\bbtheta^{\mu_{R}})\in\ccalX\times\Theta$,
whereas from Lemma \ref{lem:gm_hm} it follows that
\begin{flalign}
\bbg(\bbx^{\mu_{S}})\ge\bbg_{\mu_{S}}(\bbx^{\mu_{S}}) & \ge{\bf 0}\quad\text{and}\\
\overline{\bbf}^{\bbphi}(\bbtheta^{\mu_{R}})-\bbx^{\mu_{S}}+\mu_{R}\boldsymbol{c}_{R}{\textstyle \sqrt{N_{\bbphi}}} & \ge\overline{\bbf}_{\mu_{R}}^{\bbphi}(\bbtheta^{\mu_{R}})-\bbx^{\mu_{S}}\nonumber \\
 & \ge\mathtt{S}(\mu_{R}).
\end{flalign}
Rearranging the second inequality, we obtain 
\begin{equation}
\overline{\bbf}^{\bbphi}(\bbtheta^{\mu_{R}})-\bbx^{\mu_{S}}\ge\mathtt{S}(\mu_{R})-\mu_{R}\boldsymbol{c}_{R}{\textstyle \sqrt{N_{\bbphi}}},
\end{equation}
where the right-hand-sides are independent of the feasible point $(\bbx^{\mu_{S}},\bbtheta^{\mu_{R}})$.
The result now readily follows.
\end{IEEEproof}
It would be useful to note that if $\mathtt{S}$ is such that condition
\eqref{eq:FeasCond_2} is satisfied for \textit{all} qualifying $\mu_{R}$,
then feasibility of \eqref{eq:PFA} is ensured \textit{uniformly relative
to the choice of $\mu_{R}$ (and $\mu_{S}$)}. This means that, whenever
a solution to \eqref{eq:Zero_PFA} is determined, this solution will
automatically satisfy the original resource constraints of the initial
parameterized problem \eqref{eq:PFA}.

Another important observation is that Theorems \ref{thm:S_Feasibility}
and \ref{thm:S_Feasibility-2} are \textit{not exclusive}; in other
words, they can hold simultaneously. Indeed, the former concerns choosing
$\mu_{S}$ and $\mu_{R}$, whereas the latter concerns choosing the
slack $\mathtt{S}$, which is a function $\mu_{R}$, in a way which
is compatible with Assumption \ref{AssumptionSlack}.

As an example, one can set $\mathtt{S}(\mu_{R})\equiv\mu_{R}\boldsymbol{c}_{R}{\textstyle \sqrt{N_{\bbphi}}}$,
where $\mathtt{S}$ readily satisfies Assumption \ref{AssumptionSlack}.
However, this might not be a feasible choice in practice, since the
entries of $\boldsymbol{c}_{R}$ will probably be unknown. Still,
Theorem \ref{thm:S_Feasibility-2} provides a basic principle for
choosing $\mathtt{S}$. For instance, the choice $\mathtt{S}(\mu_{R})\equiv\boldsymbol{C}\mu_{R}{\textstyle \sqrt{N_{\bbphi}}}$
would work fine, for an appropriate constant vector $\boldsymbol{C}>{\bf 0}$,
which may chosen experimentally. This last point also highlights the
operational importance of Theorem \ref{thm:S_Feasibility-2}.

\section{\label{sec:Lagrangian-Duality}Lagrangian Duality}

A promising approach for dealing with the explicit constraints of
either problems \eqref{eq:PFA} or \eqref{eq:Zero_PFA} is by exploiting
\textit{Lagrangian Duality}, which has been proven essential and undoubtedly
important in analyzing and efficiently solving constrained convex
optimization problems; see, e.g., \cite{Boyd2004,Ruszczynski2006b,Bertsekas2009}.
Note, however, that, since both problems \eqref{eq:PFA} and \eqref{eq:Zero_PFA}
are typically nonconvex, most standard results in Lagrangian Duality
for convex optimization do not apply automatically.

Instead, our treatment will be based on recent results reported in
\cite{Eisen2019}, which in turn relies on earlier results reported
in \cite{Ribeiro2012}. In particular, the purpose of
this section is to explicitly link the smoothed surrogate \eqref{eq:Zero_PFA}
to the parameterized problem \eqref{eq:PFA}, and ultimately to the
base policy search problem \eqref{eq:Base_Problem}, \textit{in the
dual domain}, effectively characterizing the respective duality gaps.
Our results essentially provide a technically grounded path to dealing
with the constrained problem \eqref{eq:Base_Problem} in the model-free
setting, through the zeroth-order proxy \eqref{eq:Zero_PFA}.

To this end, consider the \textit{Lagrangian function} $\mathscr{L}_{\bbphi}:\bbR^{N_{S}}\times\bbR^{N_{\bbphi}}\times\bbR^{N_{\bbg}}\times\bbR^{N_{S}}\to\bbR$
defined as
\begin{equation}
\mathscr{L}_{\bbphi}(\bbx,\bbtheta,\boldsymbol{\lambda})\triangleq g^{o}(\bbx)+\left\langle \boldsymbol{\lambda}_{S},\bbg(\bbx)\right\rangle +\big\langle\boldsymbol{\lambda}_{R},\overline{\bbf}^{\bbphi}(\bbtheta)-\bbx\big\rangle,
\end{equation}
where $\boldsymbol{\lambda}\equiv(\boldsymbol{\lambda}_{S},\boldsymbol{\lambda}_{R})\in\bbR^{N_{\bbg}}\times\bbR^{N_{S}}$
are multipliers associated with the respective constraint of the \textit{primal
problem} \eqref{eq:PFA}. Then the \textit{dual function} $\mathscr{D}_{\bbphi}:\bbR^{N_{\bbg}}\times\bbR^{N_{S}}\to(-\infty,\infty]$
is defined as
\begin{equation}
\mathscr{D}_{\bbphi}(\boldsymbol{\lambda})\triangleq\sup_{(\bbx,\bbtheta)\in\ccalX\times\Theta}\mathscr{L}(\bbx,\bbtheta,\boldsymbol{\lambda}).
\end{equation}
Since it is true that $\mathscr{P}_{\bbphi}^{*}\le\mathscr{D}_{\bbphi}$
\textit{on} $\bbR_{+}^{N_{\bbg}}\times\bbR_{+}^{N_{S}}$,
it is most reasonable to consider the \textit{dual problem}
\begin{equation}
\begin{array}{rl}
\mathrm{minimize} & \mathscr{D}_{\bbphi}(\boldsymbol{\lambda})\\
\mathrm{subject\,to} & \boldsymbol{\lambda}\ge{\bf 0}
\end{array},
\end{equation}
whose optimal value
\begin{equation}
\mathscr{D}_{\bbphi}^{*}\triangleq\inf_{\boldsymbol{\lambda}\ge{\bf 0}}\mathscr{D}_{\bbphi}(\boldsymbol{\lambda})\in(-\infty,\infty]
\end{equation}
serves as the tightest over-estimate of the optimal value of \eqref{eq:PFA},
$\mathscr{P}_{\bbphi}^{*}$, when knowing only $\mathscr{D}_{\bbphi}$.

In the same fashion, for $\mu_{S}>0$ and $\mu_{R}>0$,
we define the Lagrangian function $\mathscr{L}_{\mu}:\ccalX\times\Theta\times\bbR^{N_{\bbg}}\times\bbR^{N_{S}}\to\bbR$
associated with the smoothed surrogate \eqref{eq:Zero_PFA} as\vspace{-4bp}
\begin{multline}
\quad\mathscr{L}_{\bbphi,\mu}(\bbx,\bbtheta,\boldsymbol{\lambda})\triangleq g_{\mu_{S}}^{o}(\bbx)+\left\langle \boldsymbol{\lambda}_{S},\bbg_{\mu_{S}}(\bbx)\right\rangle \quad\quad\\
+\big\langle\boldsymbol{\lambda}_{R},\overline{\bbf}_{\mu_{R}}^{\bbphi}(\bbtheta)-\bbx-\mathtt{S}(\mu_{R})\big\rangle,\,\,\,
\end{multline}
whereas the dual function $\mathscr{D}_{\bbphi,\mu}:\bbR^{N_{\bbg}}\times\bbR^{N_{S}}\to(-\infty,\infty]$
and corresponding dual infimal value\textbf{ }are
\begin{flalign}
\mathscr{D}_{\bbphi,\mu}(\boldsymbol{\lambda}) & \triangleq\sup_{(\bbx,\bbtheta)\in\ccalX\times\Theta}\mathscr{L}_{\bbphi,\mu}(\bbx,\bbtheta,\boldsymbol{\lambda})\quad\text{and}\\
\mathscr{D}_{\bbphi,\mu}^{*} & \triangleq\inf_{\boldsymbol{\lambda}\ge{\bf 0}}\mathscr{D}_{\bbphi,\mu}(\boldsymbol{\lambda})\in(-\infty,\infty],
\end{flalign}
respectively. Note that the both $\mathscr{D}_{\bbphi}$
and $\mathscr{D}_{\bbphi,\mu}$ are convex on $\bbR^{N_{\bbg}}\times\bbR^{N_{S}}$,
as pointwise suprema of affine functions. In our analysis, we will
exploit another basic assumption, as follows.

\noindent \begin{assumption}\label{AssumptionFeas}Problem \eqref{eq:PFA}
is strictly feasible.\end{assumption}

Under Assumption \ref{AssumptionFeas}, it is true that the base problem
\eqref{eq:Base_Problem} is strictly feasible as well; its feasible
set contains that of \eqref{eq:PFA}.

\vspace{-2pt}
\subsection{\label{subsec:Dual-Optimal-Values}Dual Optimal Values}

Our first task will be to explicitly relate the optimal dual values
$\mathscr{D}_{\bbphi}^{*}$ and $\mathscr{D}_{\bbphi,\mu}^{*}$.
To do so, we develop and exploit the following technical results.
\begin{lem}
\textbf{\textup{(Lagrangian Approximation)}}\label{lem:Lagrange_Approximation}
Let Assumption \ref{AssumptionMain} be in effect, and for every $\mu_{S}\ge0$
, $\mu_{R}\ge0$ and $\boldsymbol{\lambda}\ge{\bf 0}$, define the
nonnegative quantities\vspace{-3bp}
\begin{flalign}
\hspace{-1pt}\hspace{-1pt}\hspace{-1pt}\Gamma_{\mu}^{l}(\boldsymbol{\lambda}) & \triangleq\mu_{S}L_{g}^{o}{\textstyle \sqrt{N_{S}}}+\mu_{S}\big\langle\boldsymbol{\lambda}_{S},\boldsymbol{c}_{S}\big\rangle{\textstyle \sqrt{N_{S}}}\nonumber \\
 & \quad+\mu_{R}\big\langle\boldsymbol{\lambda}_{R},\boldsymbol{c}_{R}\big\rangle{\textstyle \sqrt{N_{\bbphi}}}+\big\langle\mathtt{S}(\mu_{R}),\boldsymbol{\lambda}_{R}\big\rangle,\;\text{and}\\
\hspace{-1pt}\hspace{-1pt}\hspace{-1pt}\Gamma_{\mu}^{r}(\boldsymbol{\lambda}) & \triangleq\mu_{R}\big\langle\boldsymbol{\lambda}_{R},\boldsymbol{c}_{R}\big\rangle{\textstyle \sqrt{N_{\bbphi}}}-\big\langle\mathtt{S}(\mu_{R}),\boldsymbol{\lambda}_{R}\big\rangle.
\end{flalign}
Then, for every $(\bbx,\bbtheta)\in\ccalX\times\Theta$,
it is true that
\begin{equation}
-\Gamma_{\mu}^{l}(\boldsymbol{\lambda})\le\mathscr{L}_{\bbphi,\mu}(\bbx,\bbtheta,\boldsymbol{\lambda})-\mathscr{L}_{\bbphi}(\bbx,\bbtheta,\boldsymbol{\lambda})\le\Gamma_{\mu}^{r}(\boldsymbol{\lambda}).
\end{equation}
\end{lem}
\begin{IEEEproof}[Proof of Lemma \ref{lem:Lagrange_Approximation}]
Let $\mu_{S}\ge0$ and $\mu_{R}\ge0$. Since Assumption \ref{AssumptionMain}
is in effect, Lemma \ref{lem:gm_hm} implies that, for every $(\bbx,\bbtheta,\boldsymbol{\lambda})\in\ccalX\times\Theta\times\bbR_{+}^{N_{\bbg}}\times\bbR_{+}^{N_{S}}$,
\begin{flalign}
 & \hspace{-1pt}\mathscr{L}_{\bbphi,\mu}(\bbx,\bbtheta,\boldsymbol{\lambda})\nonumber \\
 & \equiv g_{\mu_{S}}^{o}(\bbx)+\left\langle \boldsymbol{\lambda}_{S},\bbg_{\mu_{S}}(\bbx)\right\rangle +\big\langle\boldsymbol{\lambda}_{R},\overline{\bbf}_{\mu_{R}}^{\bbphi}(\bbtheta)-\bbx-\mathtt{S}(\mu_{R})\big\rangle\nonumber \\
 & \le g^{o}(\bbx)+\left\langle \boldsymbol{\lambda}_{S},\bbg(\bbx)\right\rangle +\big\langle\boldsymbol{\lambda}_{R},\overline{\bbf}^{\bbphi}(\bbtheta)-\bbx\big\rangle\nonumber \\
 & \quad+\big\langle\boldsymbol{\lambda}_{R},\mu_{R}\boldsymbol{c}_{R}{\textstyle \sqrt{N_{\bbphi}}}-\mathtt{S}(\mu_{R})\big\rangle\nonumber \\
 & \equiv\mathscr{L}_{\bbphi}(\bbx,\bbtheta,\boldsymbol{\lambda})+\Gamma_{\mu}^{r}(\boldsymbol{\lambda}).\label{eq:Right}
\end{flalign}
 By symmetry, a similar argument is possible for $\mathscr{L}_{\bbphi}$,
namely,
\begin{flalign}
 & \hspace{-1pt}\mathscr{L}_{\bbphi}(\bbx,\bbtheta,\boldsymbol{\lambda})\nonumber \\
 & \equiv g^{o}(\bbx)+\left\langle \boldsymbol{\lambda}_{S},\bbg(\bbx)\right\rangle +\big\langle\boldsymbol{\lambda}_{R},\overline{\bbf}^{\bbphi}(\bbtheta)-\bbx\big\rangle\nonumber \\
 & \le g_{\mu_{S}}^{o}(\bbx)+\left\langle \boldsymbol{\lambda}_{S},\bbg_{\mu_{S}}(\bbx)\right\rangle +\big\langle\boldsymbol{\lambda}_{R},\overline{\bbf}_{\mu_{R}}^{\bbphi}(\bbtheta)-\bbx-\mathtt{S}(\mu_{R})\big\rangle\nonumber \\
 & \quad+\mu_{S}L_{g}^{o}{\textstyle \sqrt{N_{S}}}+\big\langle\boldsymbol{\lambda}_{S},\mu_{S}\boldsymbol{c}_{S}{\textstyle \sqrt{N_{S}}}\big\rangle\nonumber \\
 & \quad\quad+\big\langle\boldsymbol{\lambda}_{R},\mu_{R}\boldsymbol{c}_{R}{\textstyle \sqrt{N_{\bbphi}}}\big\rangle+\big\langle\boldsymbol{\lambda}_{R},\mathtt{S}(\mu_{R})\big\rangle\nonumber \\
 & \equiv\mathscr{L}_{\bbphi,\mu}(\bbx,\bbtheta,\boldsymbol{\lambda})+\Gamma_{\mu}^{l}(\boldsymbol{\lambda}).\label{eq:Left}
\end{flalign}
Rearranging \eqref{eq:Right} and \eqref{eq:Left} gives the result.
\end{IEEEproof}
\begin{lem}
\textbf{\textup{(Dual Functions are Proper \& Closed)}}\label{lem:Dual_ProperClosed}
As long as $\mathscr{D}_{\bbphi}^{*}<\infty$, the dual
function $\mathscr{D}_{\bbphi}$ is proper and closed.
If, additionally, Assumption \ref{AssumptionMain} is in effect, then,
for every $\mu_{S}>0$ and $\mu_{R}>0$, the smoothed dual function
$\mathscr{D}_{\bbphi,\mu}$ is also proper and closed.
\end{lem}
\begin{IEEEproof}[Proof of Lemma \ref{lem:Dual_ProperClosed}]
Since $\mathscr{D}_{\bbphi}^{*}<\infty$, there exists
a dual feasible point $\boldsymbol{\lambda}^{\ddagger}\ge{\bf 0}$
such that $\mathscr{D}_{\bbphi}(\boldsymbol{\lambda}^{\ddagger})<\infty$,
whereas the fact that $\mathscr{L}_{\bbphi}$ is real-valued
on its domain implies that $\mathscr{D}_{\bbphi}>-\infty$
on $\bbR^{N_{\bbg}}\times\bbR^{N_{S}}$. Then
$\mathscr{D}_{\bbphi}$ is proper, by definition (p. 7
in \cite{Bertsekas2009}).

To show that $\mathscr{D}_{\bbphi}$ is also closed, it
suffices to observe that it is the pointwise supremum of affine functions,
each of which is continuous (thus lower semicontinuous) on the closed
set $\bbR^{N_{\bbg}}\times\bbR^{N_{S}}$, and,
therefore, closed (Proposition 1.1.3 in \cite{Bertsekas2009}). Then
$\mathscr{D}_{\bbphi}$ must be closed, by (\cite{Bertsekas2009},
Proposition 1.1.6).

If now Assumption \ref{AssumptionMain} is in effect, then for the
same dual feasible point $\boldsymbol{\lambda}^{\ddagger}$ as above,
and for every $(\bbx,\bbtheta)\in\ccalX\times\Theta$,
Lemma \ref{lem:Lagrange_Approximation} implies that
\begin{equation}
\mathscr{L}_{\bbphi,\mu}(\bbx,\bbtheta,\boldsymbol{\lambda}^{\ddagger})\le\mathscr{L}_{\bbphi}(\bbx,\bbtheta,\boldsymbol{\lambda}^{\ddagger})+\Gamma_{\mu}^{r}(\boldsymbol{\lambda}^{\ddagger}).
\end{equation}
Therefore, it follows that
\begin{flalign}
\mathscr{D}_{\bbphi,\mu}(\boldsymbol{\lambda}^{\ddagger}) & \equiv\sup_{(\bbx,\bbtheta)\in\ccalX\times\Theta}\mathscr{L}_{\bbphi,\mu}(\bbx,\bbtheta,\boldsymbol{\lambda}^{\ddagger})\nonumber \\
 & \le\bigg[\sup_{(\bbx,\bbtheta)\in\ccalX\times\Theta}\mathscr{L}_{\bbphi}(\bbx,\bbtheta,\boldsymbol{\lambda}^{\ddagger})\bigg]+\Gamma_{\mu}^{r}(\boldsymbol{\lambda}^{\ddagger})\nonumber \\
 & \equiv\mathscr{D}_{\bbphi}(\boldsymbol{\lambda}^{\ddagger})+\Gamma_{\mu}^{r}(\boldsymbol{\lambda}^{\ddagger})<\infty.
\end{flalign}
As before, $\mathscr{L}_{\bbphi,\mu}$ is real-valued on
its domain, thus $\mathscr{D}_{\bbphi,\mu}>-\infty$ everywhere
on $\bbR^{N_{\bbg}}\times\bbR^{N_{S}}$, showing
that $\mathscr{D}_{\bbphi,\mu}$ is proper. Lastly, closeness
of $\mathscr{D}_{\bbphi,\mu}$ follows by the same argument
as that for $\mathscr{D}_{\bbphi}$, above.
\end{IEEEproof}

\begin{lem}
\textbf{\textup{(Existence of Dual Optimal Solutions)}}\label{lem:Dual_Existence}
Suppose that $\mathscr{D}_{\bbphi}^{*}<\infty$, and let
Assumption \ref{AssumptionFeas} be in effect. Then, the set of dual
optimal solutions $\arg\min_{\boldsymbol{\lambda}\ge{\bf 0}}\mathscr{D}_{\bbphi}(\boldsymbol{\lambda})$
is nonempty and compact in $\bbR_{+}^{N_{\bbg}}\times\bbR_{+}^{N_{S}}$.
If, additionally, Assumptions \ref{AssumptionMain} and \ref{AssumptionSlack}
are in effect, then there exist $\mu_{S}^{\dagger}>0$ and $\mu_{R}^{\dagger}>0$,
such that, for every $0<\mu_{S}\le\mu_{S}^{\dagger}$ and $0<\mu_{R}\le\mu_{R}^{\dagger}$,
the solution set $\arg\min_{\boldsymbol{\lambda}\ge{\bf 0}}\mathscr{D}_{\bbphi,\mu}(\boldsymbol{\lambda})$
is a nonempty and compact subset of $\bbR_{+}^{N_{\bbg}}\times\bbR_{+}^{N_{S}}$,
as well.
\end{lem}
\begin{IEEEproof}[Proof of Lemma \ref{lem:Dual_Existence}]
For any point $(\bbx^{\dagger},\bbtheta^{\dagger})\in\bbR^{N_{S}}\times\bbR^{N_{\bbphi}}$,
which is strictly feasible for \eqref{eq:PFA}, it is true that $\bbg(\bbx^{\dagger})>{\bf 0}$
and $\overline{\bbf}^{\bbphi}(\bbtheta^{\dagger})-\bbx^{\dagger}>{\bf 0}$.
Then, for every $\boldsymbol{\lambda}\ge{\bf 0}$, we have
\begin{equation}
\mathscr{L}_{\bbphi}(\bbx^{\dagger},\bbtheta^{\dagger},\boldsymbol{\lambda})\le\mathscr{D}_{\bbphi}(\boldsymbol{\lambda}),
\end{equation}
where the left-hand-side is \textit{non-trivially affine in} $\boldsymbol{\lambda}$.
Now, consider any sequence $\{\boldsymbol{\lambda}^{n}\}_{n\in\mathbb{N}}\hspace{-1pt}\hspace{-1pt}\hspace{-0.5pt}\subseteq\hspace{-0.5pt}\hspace{-1pt}\hspace{-1pt}\bbR_{+}^{N_{\bbg}}\hspace{-1pt}\hspace{-1pt}\hspace{-0.5pt}\times\hspace{-0.5pt}\hspace{-1pt}\bbR_{+}^{N_{S}}$,
such that $\lim_{n\to\infty}\Vert\boldsymbol{\lambda}^{n}\Vert_{2}\equiv\infty$.
Since $\mathscr{L}_{\bbphi}(\bbx^{\dagger},\bbtheta^{\dagger},\cdot)$
is a non-trivial affine function \textit{and with positive slope},
we may write
\begin{flalign}
\liminf_{n\to\infty}\mathscr{D}_{\bbphi}(\boldsymbol{\lambda}^{n}) & \ge\liminf_{n\to\infty}\mathscr{L}_{\bbphi}(\bbx^{\dagger},\bbtheta^{\dagger},\boldsymbol{\lambda}^{n})\nonumber \\
 & \equiv\lim_{n\to\infty}\mathscr{L}_{\bbphi}(\bbx^{\dagger},\bbtheta^{\dagger},\boldsymbol{\lambda}^{n})\equiv\infty,
\end{flalign}
which in turn yields
\begin{equation}
\liminf_{n\to\infty}\mathscr{D}_{\bbphi}(\boldsymbol{\lambda}^{n})\equiv\infty\implies\lim_{n\to\infty}\mathscr{D}_{\bbphi}(\boldsymbol{\lambda}^{n})\equiv\infty.
\end{equation}
Noting that $\mathscr{D}_{\bbphi}$ is proper due to $\mathscr{D}_{\bbphi}^{*}$
being finite by assumption and Lemma \ref{lem:Dual_ProperClosed},
it is straightforward to show that the \textit{clipped dual function}
$\widetilde{\mathscr{D}}_{\bbphi}:\bbR^{N_{\bbg}}\times\bbR^{N_{S}}\to(-\infty,\infty]$
defined as
\begin{equation}
\widetilde{\mathscr{D}}_{\bbphi}(\boldsymbol{\lambda})\triangleq\begin{cases}
\mathscr{D}_{\bbphi}(\boldsymbol{\lambda}), & \text{if }\boldsymbol{\lambda}\ge{\bf 0}\\
\infty, & \text{otherwise}
\end{cases}
\end{equation}
is proper and \textit{coercive }(p. 119 in \cite{Bertsekas2009}).
Since, also by Lemma \ref{lem:Dual_ProperClosed}, $\mathscr{D}_{\bbphi}$
is closed, it follows that $\widetilde{\mathscr{D}}_{\bbphi}$
is closed as well (as a sum of proper closed functions: $\mathscr{D}_{\bbphi}$
itself and the indicator of $\bbR_{+}^{N_{\bbg}}\times\bbR_{+}^{N_{S}}$;
see (\cite{Bertsekas2009}, Proposition 1.1.5)). Thus, we may call
(\cite{Bertsekas2009}, Proposition 3.2.1), which ensures that $\arg\min_{\boldsymbol{\lambda}\ge{\bf 0}}\mathscr{D}_{\bbphi}(\boldsymbol{\lambda})$
is a nonempty and compact set in $\bbR_{+}^{N_{\bbg}}\times\bbR_{+}^{N_{S}}$.

Next, whenever Assumptions \ref{AssumptionMain} and \ref{AssumptionSlack}
are in effect, Theorem \ref{thm:S_Feasibility} ensures the existence
of strictly positive numbers $\mu_{S}^{\dagger}>0$ and $\mu_{R}^{\dagger}>0$,
possibly dependent on $(\bbx^{\dagger},\bbtheta^{\dagger})$,
such that, for every $0<\mu_{S}\le\mu_{S}^{\dagger}$ and $0<\mu_{R}\le\mu_{R}^{\dagger}$,
the particular point $(\bbx^{\dagger},\bbtheta^{\dagger})$
is also strictly feasible for problem \eqref{eq:Zero_PFA}. Then,
with the help of the respective part of Lemma \ref{lem:Dual_ProperClosed},
the procedure presented above for \eqref{eq:PFA} may be repeated
for \eqref{eq:Zero_PFA}, for each pair of qualifying $\mu_{S}$ and
$\mu_{R}$.
\end{IEEEproof}
Out of Lemmata \ref{lem:Lagrange_Approximation}, \ref{lem:Dual_ProperClosed}
and \ref{lem:Dual_Existence}, the first and the last are the most
important. In particular, Lemma \ref{lem:Lagrange_Approximation}
provides explicit non-symmetric upper and lower bounds for the difference
between the Lagrangians of \eqref{eq:PFA} and \eqref{eq:Zero_PFA},
\textit{which are independent of $\bbx$ and }$\bbtheta$.
Therefore, it should be possible to obtain approximation bounds on
the respective dual functions, \textit{uniform relative to $(\bbx,\bbtheta)$}.
On top of this, Lemma \ref{lem:Dual_Existence} verifies the existence
of dual optimal solutions for \eqref{eq:PFA} and \eqref{eq:Zero_PFA},
which could be exploited in conjunction with the aforementioned uniform
bounds on the respective dual functions, to provide fundamental upper
and lower bounds on the corresponding dual optimal (infimal) values.
All this is confirmed and rigorously quantified by the next central
result.
\begin{thm}
\textbf{\textup{(Dual Value Approximation)}}\label{thm:Dual_Approximation}
Let Assumptions \ref{AssumptionMain}, \ref{AssumptionSlack} and
\ref{AssumptionFeas} be in effect, and suppose that $\mathscr{D}_{\bbphi}^{*}<\infty$.
Then there exist $\mu_{S}^{\dagger}>0$ and $\mu_{R}^{\dagger}>0$,
such that, for every $0\le\mu_{S}\le\mu_{S}^{\dagger}$ and $0\le\mu_{R}\le\mu_{R}^{\dagger}$,
\begin{equation}
\boxed{-\Gamma_{\mu}^{l}(\boldsymbol{\lambda}_{\bbphi}^{\dagger})\le-\Gamma_{\mu}^{l}(\boldsymbol{\lambda}_{\bbphi,\mu}^{*})\le\mathscr{D}_{\bbphi,\mu}^{*}-\mathscr{D}_{\bbphi}^{*}\le\Gamma_{\mu}^{r}(\boldsymbol{\lambda}_{\bbphi}^{*}).}\label{eq:Main_Double}
\end{equation}
where $\boldsymbol{\lambda}_{\bbphi,\mu}^{*}\in\arg\min_{\boldsymbol{\lambda}\ge{\bf 0}}\mathscr{D}_{\bbphi,\mu}(\boldsymbol{\lambda})$,
$\mathscr{D}_{\bbphi,0}^{*}\equiv\mathscr{D}_{\bbphi}^{*}$
and $\boldsymbol{\lambda}_{\bbphi,0}^{*}\equiv\boldsymbol{\lambda}_{\bbphi}^{*}$,
identically, and where $\boldsymbol{\lambda}_{\bbphi}^{\dagger}\ge{\bf 0}$
is a finite constant, problem dependent but independent of $\mu_{S}$
and $\mu_{R}$. Further, if $\mathtt{S}(\mu_{R})\equiv\boldsymbol{C}\mu_{R}{\textstyle \sqrt{N_{\bbphi}}},\boldsymbol{C}\ge{\bf 0}$,
then there always exist finite constants $\Sigma_{S}^{l}\ge0$, $\Sigma_{R}^{l}\ge0$
and $\Sigma_{R}^{r}\in\bbR$, problem dependent but independent
of $\mu_{S}$ and $\mu_{R}$, such that
\begin{equation}
\boxed{-(\mu_{S}\Sigma_{S}^{l}+\mu_{R}\Sigma_{R}^{l})\le\mathscr{D}_{\bbphi,\mu}^{*}-\mathscr{D}_{\bbphi}^{*}\le\mu_{R}\Sigma_{R}^{r}.}\label{eq:Main_Double-2}
\end{equation}
In particular, $\Sigma_{S}^{l}$, $\Sigma_{R}^{l}$ and $\Sigma^{r}$
are defined as
\begin{flalign}
\Sigma_{S}^{l} & \triangleq\big(L_{g}^{o}+\big\langle\boldsymbol{\lambda}_{\bbphi,S}^{\dagger},\boldsymbol{c}_{S}\big\rangle\big){\textstyle \sqrt{N_{S}}},\\
\Sigma_{R}^{l} & \triangleq\big\langle\boldsymbol{\lambda}_{\bbphi,R}^{\dagger},\boldsymbol{c}_{R}+\boldsymbol{C}\big\rangle{\textstyle \sqrt{N_{\bbphi}}}\quad\text{and}\\
\Sigma_{R}^{r} & \triangleq\big\langle\boldsymbol{\lambda}_{\bbphi,R}^{*},\boldsymbol{c}_{R}-\boldsymbol{C}\big\rangle{\textstyle \sqrt{N_{\bbphi}}}.
\end{flalign}
Lastly, whenever $\mathtt{S}(\mu_{R})\equiv\boldsymbol{C}\mu_{R}{\textstyle \sqrt{N_{\bbphi}}}$
with $\boldsymbol{C}\ge\boldsymbol{c}_{R}$, then the right-hand-sides
of \eqref{eq:Main_Double} and \eqref{eq:Main_Double-2} are nonpositive,
and may be replaced by zero.
\end{thm}
\begin{IEEEproof}[Proof of Theorem \ref{thm:Dual_Approximation}]
Under the assumptions of the theorem, Lemma \ref{lem:Dual_Existence}
ensures that the dual optimal solution set $\arg\min_{\boldsymbol{\lambda}\ge{\bf 0}}\mathscr{D}_{\bbphi,\mu}(\boldsymbol{\lambda})$
is nonempty, for all $0\le\mu_{S}\le\mu_{S}^{\dagger}$ and $0\le\mu_{R}\le\mu_{R}^{\dagger}$.
Consequently, there exist optimal multipliers $\boldsymbol{\lambda}_{\bbphi,\mu}^{*}\ge{\bf 0}$,
such that
\begin{equation}
-\infty<\mathscr{D}_{\bbphi,\mu}(\boldsymbol{\lambda}_{\bbphi,\mu}^{*})\equiv\mathscr{D}_{\bbphi,\mu}^{*}\equiv\inf_{\boldsymbol{\lambda\ge{\bf 0}}}\mathscr{D}_{\bbphi,\mu}(\boldsymbol{\lambda}),
\end{equation}
for all allowable values of $\mu_{S}$ and $\mu_{R}$. Therefore,
invoking Lemma \ref{lem:Lagrange_Approximation}, we may carefully
write
\begin{flalign}
\mathscr{D}_{\bbphi,\mu}^{*} & \le\mathscr{D}_{\bbphi,\mu}(\boldsymbol{\lambda}_{\bbphi}^{*})\nonumber \\
 & \equiv\sup_{(\bbx,\bbtheta)\in\ccalX\times\Theta}\mathscr{L}_{\bbphi,\mu}(\bbx,\bbtheta,\boldsymbol{\lambda}_{\bbphi}^{*})\nonumber \\
 & \le\bigg[\sup_{(\bbx,\bbtheta)\in\ccalX\times\Theta}\mathscr{L}_{\bbphi}(\bbx,\bbtheta,\boldsymbol{\lambda}_{\bbphi}^{*})\bigg]+\Gamma_{\mu}^{r}(\boldsymbol{\lambda}_{\bbphi}^{*})\nonumber \\
 & \equiv\mathscr{D}_{\bbphi}(\boldsymbol{\lambda}_{\bbphi}^{*})+\Gamma_{\mu}^{r}(\boldsymbol{\lambda}_{\bbphi}^{*})\nonumber \\
 & \equiv\mathscr{D}_{\bbphi}^{*}+\Gamma_{\mu}^{r}(\boldsymbol{\lambda}_{\bbphi}^{*}).\label{eq:Main_Right}
\end{flalign}
By symmetry, we also have
\begin{flalign}
\mathscr{D}_{\bbphi}^{*} & \le\mathscr{D}_{\bbphi}(\boldsymbol{\lambda}_{\bbphi,\mu}^{*})\nonumber \\
 & \equiv\sup_{(\bbx,\bbtheta)\in\ccalX\times\Theta}\mathscr{L}_{\bbphi}(\bbx,\bbtheta,\boldsymbol{\lambda}_{\bbphi,\mu}^{*})\nonumber \\
 & \le\bigg[\sup_{(\bbx,\bbtheta)\in\ccalX\times\Theta}\mathscr{L}_{\bbphi,\mu}(\bbx,\bbtheta,\boldsymbol{\lambda}_{\bbphi,\mu}^{*})\bigg]+\Gamma_{\mu}^{l}(\boldsymbol{\lambda}_{\bbphi,\mu}^{*})\nonumber \\
 & \equiv\mathscr{D}_{\bbphi,\mu}(\boldsymbol{\lambda}_{\bbphi,\mu}^{*})+\Gamma_{\mu}^{l}(\boldsymbol{\lambda}_{\bbphi,\mu}^{*})\nonumber \\
 & \equiv\mathscr{D}_{\bbphi}^{*}+\Gamma_{\mu}^{l}(\boldsymbol{\lambda}_{\bbphi,\mu}^{*}).\label{eq:Main_Left}
\end{flalign}
Rearranging \eqref{eq:Main_Right} and \eqref{eq:Main_Left}, we obtain
the last three inequalities of \eqref{eq:Main_Double} (left-to-right),
as in the statement of the theorem.

Now, recalling that $\boldsymbol{\lambda}_{\bbphi,\mu}^{*}\equiv(\boldsymbol{\lambda}_{\bbphi,S}^{*}(\mu_{S}),\boldsymbol{\lambda}_{\bbphi,R}^{*}(\mu_{R}))$,
we may define \textit{maximal multipliers }(here the supremum is taken
\textit{elementwise} on the involved vectors)
\begin{flalign}
\boldsymbol{\lambda}_{\bbphi,S}^{\dagger} & \triangleq\sup_{\mu\in[0,\mu_{S}^{\dagger}]}\boldsymbol{\lambda}_{\bbphi,S}^{*}(\mu_{S}),\quad\text{and}\\
\boldsymbol{\lambda}_{\bbphi,R}^{\dagger} & \triangleq\sup_{\mu\in[0,\mu_{R}^{\dagger}]}\boldsymbol{\lambda}_{\bbphi,R}^{*}(\mu_{R}).
\end{flalign}
Note that $\boldsymbol{\lambda}_{\bbphi}^{\dagger}\equiv(\boldsymbol{\lambda}_{\bbphi,S}^{\dagger},\boldsymbol{\lambda}_{\bbphi,R}^{\dagger})\ge{\bf 0}$
is finite, since $\boldsymbol{\lambda}_{\bbphi,S}^{*}(\mu_{S})\ge{\bf 0}$
and $\boldsymbol{\lambda}_{\bbphi,R}^{*}(\mu_{R})\ge{\bf 0}$
are finite everywhere on the compact sets $[0,\mu_{S}^{\dagger}]$
and $[0,\mu_{R}^{\dagger}]$, respectively. Using these definitions,
it follows that, for every $0\le\mu_{S}\le\mu_{S}^{\dagger}$ and
$0\le\mu_{R}\le\mu_{R}^{\dagger}$,
\begin{equation}
\Gamma_{\mu}^{l}(\boldsymbol{\lambda}_{\bbphi,\mu}^{*})\le\Gamma_{\mu}^{l}(\boldsymbol{\lambda}_{\bbphi}^{\dagger}),
\end{equation}
verifying the left inequality of \eqref{eq:Main_Double}.

When $\mathtt{S}(\mu_{R})\equiv\boldsymbol{C}\mu_{R}{\textstyle \sqrt{N_{\bbphi}}},\boldsymbol{C}\ge{\bf 0}$,
the rest of the claims stated in the theorem follow by noting that,
for every $\boldsymbol{\lambda}\ge{\bf 0}$,
\begin{flalign}
\hspace{-1pt}\hspace{-1pt}\hspace{-1pt}\hspace{-1pt}\Gamma_{\mu}^{l}(\boldsymbol{\lambda}) & \triangleq\mu_{S}\big(L_{g}^{o}+\big\langle\boldsymbol{\lambda}_{S},\boldsymbol{c}_{S}\big\rangle\hspace{-1pt}\big){\textstyle \sqrt{N_{S}}}\nonumber \\
 & \quad+\mu_{R}\big(\hspace{-1pt}\big\langle\boldsymbol{\lambda}_{R},\boldsymbol{c}_{R}\big\rangle+\big\langle\boldsymbol{C},\boldsymbol{\lambda}_{R}\big\rangle\hspace{-1pt}\big){\textstyle \sqrt{N_{\bbphi}}}\quad\text{and}\\
\hspace{-1pt}\hspace{-1pt}\hspace{-1pt}\hspace{-1pt}\Gamma_{\mu}^{r}(\boldsymbol{\lambda}) & \triangleq\mu_{R}\big(\big\langle\boldsymbol{\lambda}_{R},\boldsymbol{c}_{R}\big\rangle-\big\langle\boldsymbol{C},\boldsymbol{\lambda}_{R}\big\rangle\big){\textstyle \sqrt{N_{\bbphi}}},\nonumber 
\end{flalign}
where both functions $\Gamma_{(\cdot)}^{l}$ and $\Gamma_{(\cdot)}^{r}$
are nondecreasing in both $\mu_{S}$ and $\mu_{R}$. If, additionally,
$\boldsymbol{C}\ge\boldsymbol{c}_{R}$, then $\Gamma_{(\cdot)}^{r}\le0$.
The proof is complete.
\end{IEEEproof}

\subsection{\label{subsec:Strong-Duality}Approximate Strong Duality}

After explicitly relating the dual optimal values $\mathscr{D}_{\bbphi}^{*}$
and $\mathscr{D}_{\bbphi,\mu}^{*}$, our second task will
be to relate $\mathscr{D}_{\bbphi,\mu}^{*}$ to the optimal
value of the base problem \eqref{eq:Base_Problem}. In particular,
we would like to characterize the \textit{duality gap between the
primal problem \eqref{eq:Base_Problem} and the dual to problem \eqref{eq:Zero_PFA}}.
Note that we are \textit{not} interested in characterizing the duality
gap of \eqref{eq:Zero_PFA}; quite interestingly, to the best of our
knowledge, this constitutes a nontrivial problem.

Following \cite{Eisen2019}, we exploit the notion of a $\epsilon$-universal
policy parameterization. This allows us to characterize the \textit{intermediate}
duality gap between the optimal value of \eqref{eq:Base_Problem},
and the optimal value of the dualization of \eqref{eq:Zero_PFA} \cite{Eisen2019}.
\begin{defn}
\textbf{($\epsilon$-Universality)\label{def:Universal-100}} Fix
$\epsilon\ge0$, choose a parameterization $\bbphi\hspace{-1pt}:\hspace{-1pt}\ccalH\times\bbR^{N_{\bbphi}}\hspace{-1pt}\hspace{-1pt}\to\hspace{-1pt}\bbR^{N_{R}}$,
and let $\Theta\hspace{-1pt}\subseteq\hspace{-0.5pt}\bbR^{N_{\bbphi}}$
be any parameter subspace. A class of admissible policies
\begin{equation}
\ccalP _{\Theta}^{\bbphi}\triangleq\{\bbp\in\ccalP \,\hspace{-0.5pt}|\,\bbp(\cdot)\equiv\bbphi(\cdot,\bbtheta),\;\bbtheta\in\Theta\}\subseteq\ccalP 
\end{equation}
is called \textit{$\epsilon$-universal in $\ccalP $} if and only
if, for every $\bbp\in\ccalP $, there exists $\bbphi(\cdot,\bbtheta\equiv\bbtheta(\epsilon,\bbp))\in\ccalP _{\Theta}^{\bbphi}$,
such that
\begin{equation}
\mbE\{\Vert\bbp(\bbH)-\bbphi(\bbH,\bbtheta)\Vert_{\infty}\}<\epsilon.\label{eq:e_universal100}
\end{equation}
\end{defn}
\begin{rem}
Note that, when $\Theta\subseteq\big\{\bbtheta\in\bbR^{N_{\bbphi}}\big|\hspace{0.5bp}\bbphi(\cdot,\bbtheta)\hspace{-1pt}\in\hspace{-1pt}\ccalP \big\}$,
as assumed in Section \ref{sec:Optimal-Resource-Allocation}, it follows
that $\ccalP _{\Theta}^{\bbphi}\hspace{-1pt}\equiv\hspace{-1pt}\hspace{-1pt}\{\bbphi(\cdot,\bbtheta)|\hspace{0.5bp}\bbtheta\in\Theta\}$.
Then, trivially, $\bbphi(\cdot,\bbtheta)\hspace{-1pt}\in\hspace{-1pt}\ccalP _{\Theta}^{\bbphi}$,
for all $\bbtheta\in\Theta$, and $\epsilon$-universality
of $\ccalP _{\Theta}^{\bbphi}$ in $\ccalP $ is ensured
as long as, for every admissible policy $\bbp\in\ccalP $,
there is at least one $\bbtheta\in\Theta$ satisfying \eqref{eq:e_universal100}.\hfill{}\Halmos

Additionally, also as in \cite{Eisen2019}, we will impose the following
additional structural assumptions.
\end{rem}
\begin{assumption}\label{AssumptionNonatomic}The Borel pushforward
${\cal M}_{\bbH}:\mathpzc{B}(\ccalH)\to[0,1]$
is nonatomic: For every Borel set ${\cal E}\in\mathpzc{B}(\ccalH)$
such that ${\cal M}_{\bbH}({\cal E})>0$, there exists another
Borel set ${\cal E}^{o}\in\mathpzc{B}(\ccalH)$ satisfying ${\cal M}_{\bbH}({\cal E})>{\cal M}_{\bbH}({\cal E}^{o})>0$.\end{assumption}\vspace{-4bp}

\noindent \begin{assumption}\label{AssumptionIncreasing}For every
pair $(\bbx,\bbx')\in\ccalX\times\ccalX$
such that $\bbx\le\bbx'$, it is true that $g^{o}(\bbx)\le g^{o}(\bbx')$
and $\bbg(\bbx)\le\bbg(\bbx')$.\end{assumption}\vspace{-4bp}

\noindent \begin{assumption}\label{AssumptionLip}There exists a
number $L_{\ccalP }^{\bbf}<\infty$, such that, for every
pair $(\bbp,\bbp')\in\ccalP \times\ccalP $,
it is true that\vspace{-3bp}
\begin{multline}
\quad\quad\;\;\Vert\mbE\{\bbf(\bbp(\bbH),\bbH)\}-\mbE\{\bbf(\bbp'(\bbH),\bbH)\}\Vert_{\infty}\\
\le L_{\ccalP }^{\bbf}\mbE\{\Vert\bbp(\bbH)-\bbp'(\bbH)\Vert_{\infty}\}.\quad\quad\quad\quad\;\;\;\;
\end{multline}
\end{assumption}

As clearly explained in (\cite{Eisen2019}, Section III.A), Assumptions
\ref{AssumptionNonatomic}, \ref{AssumptionIncreasing} and \ref{AssumptionLip}
are reasonable and are fulfilled by most practically significant wireless
resource allocation problems. We thus do not further comment.

Under Assumptions \ref{AssumptionFeas}, \ref{AssumptionNonatomic},
\ref{AssumptionIncreasing} and \ref{AssumptionLip}, an important
result was presented in \cite{Eisen2019}, which characterizes the
duality gap between the base problem \eqref{eq:Base_Problem} and
the parameterized surrogate \eqref{eq:PFA}, leveraging the notion
of $\epsilon$-universality of Definition \ref{def:Universal-100}.
For completeness, we also report this result here, as follows.
\begin{thm}
\textbf{\textup{(PFA Duality Gap \cite{Eisen2019})}}\label{thm:Dual_Gap_1}
Let Assumptions \ref{AssumptionFeas}, \ref{AssumptionNonatomic},
\ref{AssumptionIncreasing} and \ref{AssumptionLip} be in effect,
and suppose that, for some $\epsilon\ge0$, $\bbphi$ is
$\epsilon$-universal in $\ccalP $. Then it is true that $\mathscr{D}_{\bbphi}^{*}<\infty$
and, further,
\begin{equation}
-\Vert\boldsymbol{\lambda}^{*}\Vert L_{\ccalP }^{\bbf}\epsilon\le\mathscr{D}_{\bbphi}^{*}-\mathscr{P}^{*}\le0,
\end{equation}
where $\boldsymbol{\lambda}^{*}\in\arg\min_{\boldsymbol{\lambda}\ge{\bf 0}}\mathscr{D}(\boldsymbol{\lambda})\neq\emptyset$,
and $\mathscr{D}:\bbR^{N_{\bbg}}\times\bbR^{N_{S}}\to(-\infty,\infty]$
denotes the dual function of the base problem \eqref{eq:Base_Problem}.
\end{thm}
We now combine Theorem \ref{thm:Dual_Gap_1} with Theorem \ref{thm:Dual_Approximation}
developed in Section \ref{subsec:Dual-Optimal-Values}, resulting
in the main result of this paper. The proof is elementary, and thus
omitted.
\begin{thm}
\textbf{\textup{(Smoothed PFA Duality Gap)}}\label{thm:Dual_Gap_Smoothed}
Let Assumptions \ref{AssumptionMain}, \ref{AssumptionSlack}, \ref{AssumptionFeas},
\ref{AssumptionNonatomic}, \ref{AssumptionIncreasing} and \ref{AssumptionLip}
be in effect, and suppose that, for some $\epsilon\ge0$, $\bbphi$
is $\epsilon$-universal in $\ccalP $. Then, by the definitions
of Lemma \ref{lem:Lagrange_Approximation} and Theorem \ref{thm:Dual_Approximation},
there exist $\mu_{S}^{\dagger}>0$ and $\mu_{R}^{\dagger}>0$, such
that, for every $0\le\mu_{S}\le\mu_{S}^{\dagger}$ and $0\le\mu_{R}\le\mu_{R}^{\dagger}$,
\begin{equation}
\boxed{\hspace{-1pt}-\big(\Gamma_{\mu}^{l}(\boldsymbol{\lambda}_{\bbphi,\mu}^{*})\hspace{-1pt}+\hspace{-1pt}\Vert\boldsymbol{\lambda}^{*}\Vert L_{\ccalP }^{\bbf}\epsilon\big)\hspace{-1pt}\le\hspace{-1pt}\mathscr{D}_{\bbphi,\mu}^{*}\hspace{-1pt}-\hspace{-1pt}\mathscr{P}^{*}\hspace{-1pt}\hspace{-1pt}\le\hspace{-1pt}\Gamma_{\mu}^{r}(\boldsymbol{\lambda}_{\bbphi}^{*}),\hspace{-1pt}}\label{eq:Main2_1}
\end{equation}
with $\Gamma_{\mu}^{l}(\boldsymbol{\lambda}_{\bbphi,\mu}^{*})\le\Gamma_{\mu}^{l}(\boldsymbol{\lambda}_{\bbphi}^{\dagger})$.
Further, if $\mathtt{S}(\mu_{R})\equiv\boldsymbol{C}\mu_{R},\boldsymbol{C}\ge{\bf 0}$,
it is true that
\begin{equation}
\hspace{-1pt}\hspace{-1pt}\boxed{\hspace{-1pt}-(\mu_{S}\Sigma_{S}^{l}\hspace{-1pt}+\hspace{-1pt}\mu_{R}\Sigma_{R}^{l}\hspace{-1pt}+\hspace{-1pt}\Vert\boldsymbol{\lambda}^{*}\Vert L_{\ccalP }^{\bbf}\epsilon)\hspace{-1pt}\le\hspace{-1pt}\mathscr{D}_{\bbphi,\mu}^{*}\hspace{-1pt}-\hspace{-1pt}\mathscr{P}^{*}\hspace{-1pt}\hspace{-1pt}\le\hspace{-1pt}\mu_{R}\Sigma_{R}^{r}.\hspace{-1pt}}\hspace{-1pt}\hspace{-1pt}\label{eq:Main2_2}
\end{equation}
Lastly, whenever $\mathtt{S}(\mu_{R})\equiv\boldsymbol{C}\mu_{R}{\textstyle \sqrt{N_{\bbphi}}}$
with $\boldsymbol{C}\ge\boldsymbol{c}_{R}$, then the right-hand-sides
of \eqref{eq:Main2_1} and \eqref{eq:Main2_2} are nonpositive, and
may be replaced by zero.
\end{thm}
We may also state a trivial corollary to Theorem \ref{thm:Dual_Gap_Smoothed},
masking all its technicalities, which sometimes might be unnecessary
in more qualitative arguments.
\begin{cor}
\textbf{\textup{(Smoothed PFA Duality Gap | Simplified)}}\label{cor:Dual_Gap_Smoothed-Simple}
Let Assumptions \ref{AssumptionMain}, \ref{AssumptionSlack}, \ref{AssumptionFeas},
\ref{AssumptionNonatomic}, \ref{AssumptionIncreasing} and \ref{AssumptionLip}
be in effect and suppose that, for some $\epsilon\ge0$, $\bbphi$
is $\epsilon$-universal in $\ccalP $. Further, choose $\mathtt{S}(\mu_{R})\equiv\boldsymbol{C}\mu_{R}{\textstyle \sqrt{N_{\bbphi}}},\boldsymbol{C}\ge{\bf 0}$.
Then, it is true that
\begin{equation}
\boxed{\big|\mathscr{D}_{\bbphi,\mu}^{*}-\mathscr{P}^{*}\hspace{-0.5pt}\big|\equiv{\cal O}(\mu_{S}{\textstyle \sqrt{N_{S}}}+\mu_{R}{\textstyle \sqrt{N_{\bbphi}}}+\epsilon),}\label{eq:Main2_1-1}
\end{equation}
as $(\mu_{S},\mu_{R},\epsilon)\hspace{-0.5pt}\to\hspace{-0.5pt}{\bf 0}$
and $(N_{S},N_{\bbphi})\hspace{-0.5pt}\to\hspace{-0.5pt}\infty$.
If, further, $\boldsymbol{C}\ge\boldsymbol{c}_{R}$, \eqref{eq:Main2_1-1}
may be improved as
\begin{equation}
\boxed{0\le\mathscr{P}^{*}-\mathscr{D}_{\bbphi,\mu}^{*}\equiv{\cal O}(\mu_{S}{\textstyle \sqrt{N_{S}}}+\mu_{R}{\textstyle \sqrt{N_{\bbphi}}}+\epsilon),}\label{eq:Main2_2-1}
\end{equation}
as $(\mu_{S},\mu_{R},\epsilon)\hspace{-0.5pt}\to\hspace{-0.5pt}{\bf 0}$
and $(N_{S},N_{\bbphi})\hspace{-0.5pt}\to\hspace{-0.5pt}\infty$.
\end{cor}
Theorem \ref{thm:Dual_Gap_Smoothed} and Corollary \ref{cor:Dual_Gap_Smoothed-Simple}
explicitly quantify the gap between dual optimal value of the smoothed
surrogate \eqref{eq:Zero_PFA} and the (primal) optimal value of the
constrained variational problem \eqref{eq:Base_Problem}. What is
more, the gap can be made \textit{arbitrarily small at will}, and
scales \textit{linearly} relative to the near-universality precision
$\epsilon$, \textit{and} the smoothing parameters $\mu_{S}$ and
$\mu_{R}$.

The importance of Theorem \ref{thm:Dual_Gap_Smoothed} and Corollary
\ref{cor:Dual_Gap_Smoothed-Simple} is \textit{twofold}. First, similarly
to \cite{Eisen2019} and together with Theorems \ref{thm:S_Feasibility}
and \ref{thm:S_Feasibility-2}, Theorem \ref{thm:Dual_Gap_Smoothed}
and Corollary \ref{cor:Dual_Gap_Smoothed-Simple} provide solid technical
evidence justifying the dualization of \eqref{eq:Zero_PFA} as a proxy
for obtaining the optimal value of \eqref{eq:Base_Problem}. This
very useful per se, since the dual problem embeds the constraints
of \eqref{eq:Zero_PFA} in its objective, via the Lagrangian formulation.

Second, \textit{and most importantly}, solving for \eqref{eq:Zero_PFA}
in the dual domain can be performed in a \textit{gradient-free fashion},
using \textit{only} evaluations of the functions present in \textit{both}
the objective and constraints of \eqref{eq:Zero_PFA}, as we discuss
next. This makes optimal wireless resource allocation in the model-free
setting possible, within a \textit{non-heuristic} and \textit{predictable}
framework.

\section{\label{sec:Primal-Dual-Model-Free-Learning}Primal-Dual Model-Free
Learning}

\begin{algorithm}
	\textbf{Input:} $\bbx^{0}$, $\bbtheta^{0}$,
	$\boldsymbol{\lambda}_{S}^{0}$, $\boldsymbol{\lambda}_{R}^{0}$,
	$\{\gamma_{\bbx}^{n},\gamma_{\bbtheta}^{n},\gamma_{\boldsymbol{\lambda}_{S}}^{n},\gamma_{\boldsymbol{\lambda}_{R}}^{n}\}_{n\in\mathbb{N}}$,
	$\mu_{S}$, $\mu_{R}$
	
	\textbf{Output:} $\{\bbx^{n+1},\bbtheta^{n+1}\}_{n\in\mathbb{N}}$
	
	1:$\;$\textbf{for} $n=0,1,2,\ldots$ \textbf{do}
	
	2:$\;$$\;\,$Draw samples $\boldsymbol{U}_{S}^{n+1}$ and $\boldsymbol{U}_{R}^{n+1}$.
	
	3:$\;$$\;\,$Sample values
	\begin{align*}
	g^{o}(\bbx^{n}),\; & g^{o}(\bbx^{n}\hspace{-1pt}+\hspace{-1pt}\mu_{S}\boldsymbol{U}_{S}^{n+1})\quad\text{and}\\
	\bbg(\bbx^{n}),\; & \bbg(\bbx^{n}\hspace{-1pt}+\hspace{-1pt}\mu_{S}\boldsymbol{U}_{S}^{n+1}),
	\end{align*}
	\hphantom{3:}$\;$$\;\,$and probe the wireless system to obtain
	\begin{flalign*}
	& \bbf(\bbphi(\bbH^{n+1},\bbtheta^{n}),\bbH^{n+1})\quad\text{and}\\
	& \bbf(\bbphi(\bbH^{n+1},\bbtheta^{n}\hspace{-1pt}\hspace{-1pt}+\hspace{-1pt}\mu_{R}\boldsymbol{U}_{R}^{n+1}),\bbH^{n+1}).
	\end{flalign*}
	
	4:$\;$$\;\,$Compute $\bbx^{n+1}$ and $\bbtheta^{n+1}$
	from \eqref{eq:PrimalDual_1-1} and \eqref{eq:PrimalDual_2-1}.
	
	5:$\;$$\;\,$Sample $\bbg(\bbx^{n+1}\hspace{-1pt}+\hspace{-1pt}\mu_{S}\boldsymbol{U}_{S}^{n+1})$
	and probe the wireless system \hphantom{3:}$\;$$\;\,$to obtain
	$\bbf(\bbphi(\bbH^{n+1},\bbtheta^{n+1}\hspace{-1pt}\hspace{-1pt}+\hspace{-1pt}\mu_{R}\boldsymbol{U}_{R}^{n+1})$.
	
	6:$\;$$\;\,$Compute $\boldsymbol{\lambda}_{S}^{n+1}$and $\boldsymbol{\lambda}_{R}^{n+1}$
	from \eqref{eq:PrimalDual_4-1} and \eqref{eq:PrimalDual_4-2}.
	
	7:$\;$\textbf{end for}
	
	\caption{\label{alg:Primal-Dual}Model-Free Randomized Primal-Dual Learning}
\end{algorithm}

We now present a simple and efficient \textit{zeroth-order} randomized
primal-dual algorithm for dealing directly with the smoothed surrogate
\eqref{eq:Zero_PFA} in the model-free setting. The algorithm is \textit{non-heuristic
}and\textit{ derived} from first principles, and uses stochastic approximation
to tackle the minimax problem\vspace{-3bp}
\begin{equation}
\begin{array}{rl}
\mathrm{minimize} & \sup_{(\bbx,\bbtheta)\in\ccalX\times\Theta}\mathscr{L}_{\bbphi,\mu}(\bbx,\bbtheta,\boldsymbol{\lambda})\\
\mathrm{subject\,to} & \boldsymbol{\lambda}\ge{\bf 0}
\end{array},\label{eq:PrimalDual_1}
\end{equation}
for every qualifying choice of $\mu_{S}$ and $\mu_{R}$.

Apparently, a basic primal-dual method for solving \eqref{eq:PrimalDual_1}
may be easily derived by taking gradients relative to all of its variables,
and then performing alternating gradient steps in appropriate directions.
Specifically, for every $(\bbx,\bbtheta)\in\ccalX\times\Theta$
and for every $\boldsymbol{\lambda}\equiv(\boldsymbol{\lambda}_{S},\boldsymbol{\lambda}_{R})\ge{\bf 0}$,
the gradients of $\mathscr{L}_{\bbphi,\mu}$ with respect
to each of its arguments may be readily expressed as
\begin{flalign}
\nabla_{\bbx}\mathscr{L}_{\bbphi,\mu}(\bbx,\bbtheta,\boldsymbol{\lambda}) & \equiv\nabla g_{\mu_{S}}^{o}(\bbx)+\nabla\bbg_{\mu_{S}}(\bbx)\boldsymbol{\lambda}_{S}-\boldsymbol{\lambda}_{R},\\
\nabla_{\bbtheta}\mathscr{L}_{\bbphi,\mu}(\bbx,\bbtheta,\boldsymbol{\lambda}) & \equiv\nabla\overline{\bbf}_{\mu_{R}}^{\bbphi}(\bbtheta)\boldsymbol{\lambda}_{R},\\
\nabla_{\boldsymbol{\lambda}_{S}}\mathscr{L}_{\bbphi,\mu}(\bbx,\bbtheta,\boldsymbol{\lambda}) & \equiv\bbg_{\mu_{S}}(\bbx)\quad\text{and}\\
\nabla_{\boldsymbol{\lambda}_{R}}\mathscr{L}_{\bbphi,\mu}(\bbx,\bbtheta,\boldsymbol{\lambda}) & \equiv\overline{\bbf}_{\mu_{R}}^{\bbphi}(\bbtheta)-\bbx-\mathtt{S}(\mu_{R}).
\end{flalign}
Then, the idea is to iteratively ascend in $(\bbx,\bbtheta)$
and descend in $\boldsymbol{\lambda}$, in an alternating fashion.
If $n\in\mathbb{N}$ denotes an iteration index, this implies the
updates
\begin{flalign}
\hspace{-1pt}\hspace{-1pt}\hspace{-1pt}\hspace{-1pt}\hspace{-1pt}\bbx^{n+1} & \hspace{-1pt}\hspace{-1pt}\equiv\hspace{-0.5pt}\hspace{-1pt}\Pi_{\ccalX}\big\{\bbx^{n}\hspace{-1pt}\hspace{-1pt}+\hspace{-0.5pt}\hspace{-1pt}\boldsymbol{\gamma}_{\bbx}^{n}\circ(\nabla g_{\mu_{S}}^{o}(\bbx^{n})\nonumber \\
 & \;\;+\hspace{-1pt}\hspace{-1pt}\nabla\bbg_{\mu_{S}}(\bbx^{n})\boldsymbol{\lambda}_{S}^{n}\hspace{-1pt}\hspace{-1pt}-\hspace{-1pt}\hspace{-1pt}\boldsymbol{\lambda}_{R}^{n})\hspace{-1pt}\big\},\label{eq:PrimalDual_2}\\
\hspace{-1pt}\hspace{-1pt}\hspace{-1pt}\hspace{-1pt}\hspace{-1pt}\bbtheta^{n+1} & \hspace{-1pt}\hspace{-1pt}\equiv\hspace{-0.5pt}\hspace{-1pt}\Pi_{\Theta}\big\{\bbtheta^{n}\hspace{-1pt}\hspace{-1pt}+\hspace{-1pt}\boldsymbol{\gamma}_{\bbtheta}^{n}\circ\nabla\overline{\bbf}_{\mu_{R}}^{\bbphi}(\bbtheta^{n})\boldsymbol{\lambda}_{R}^{n}\big\},\label{eq:PrimalDual_3}\\
\hspace{-1pt}\hspace{-1pt}\hspace{-1pt}\hspace{-1pt}\hspace{-1pt}\boldsymbol{\lambda}_{S}^{n+1} & \hspace{-1pt}\hspace{-1pt}\equiv\hspace{-0.5pt}\hspace{-1pt}(\boldsymbol{\lambda}_{S}^{n}\hspace{-1pt}-\hspace{-1pt}\boldsymbol{\gamma}_{\boldsymbol{\lambda}_{S}}^{n}\hspace{-1pt}\hspace{-0.5pt}\circ\hspace{0.5pt}\bbg_{\mu_{S}}(\bbx^{n+1}))_{+}\quad\text{and}\label{eq:PrimalDual_4}\\
\hspace{-1pt}\hspace{-1pt}\hspace{-1pt}\hspace{-1pt}\hspace{-1pt}\boldsymbol{\lambda}_{R}^{n+1} & \hspace{-1pt}\hspace{-1pt}\equiv\hspace{-0.5pt}\hspace{-1pt}\big(\boldsymbol{\lambda}_{R}^{n}\hspace{-1pt}-\hspace{-1pt}\boldsymbol{\gamma}_{\boldsymbol{\lambda}_{S}}^{n}\hspace{-1pt}\hspace{-1pt}\circ\hspace{-1pt}\big(\overline{\bbf}_{\mu_{R}}^{\bbphi}(\bbtheta^{n+1})\hspace{-1pt}-\hspace{-1pt}\bbx^{n+1}\hspace{-1pt}-\hspace{-1pt}\mathtt{S}(\mu_{R})\big)\hspace{-0.5pt}\big)_{+},\hspace{-1pt}\hspace{-1pt}\hspace{-1pt}\hspace{-1pt}\label{eq:PrimalDual-5}
\end{flalign}
where ``$\circ$'' denotes the Hadamard product operation, $\{\boldsymbol{\gamma}_{\bbx}^{n}\}_{n\in\mathbb{N}}$,
$\{\boldsymbol{\gamma}_{\bbtheta}^{n}\}_{n\in\mathbb{N}}$,
$\{\boldsymbol{\gamma}_{\boldsymbol{\lambda}_{S}}^{n}\}_{n\in\mathbb{N}}$
and $\{\boldsymbol{\gamma}_{\boldsymbol{\lambda}_{R}}^{n}\}_{n\in\mathbb{N}}$
are nonnegative vector stepsize sequences, $\nabla\bbg_{\mu_{S}}:\bbR^{N_{S}}\to\bbR^{N_{S}}\times\bbR^{N_{\bbg}}$
and $\nabla\overline{\bbf}_{\mu_{R}}^{\bbphi}:\bbR^{N_{\bbphi}}\to\bbR^{N_{\bbphi}}\times\bbR^{N_{S}}$
are corresponding Jacobians and, for every nonempty closed set ${\cal A}\subseteq\bbR^{N}$,
$\Pi_{{\cal A}}:\bbR^{N}\to{\cal A}$ is the usual Euclidean
projection operator.

We may observe that the algorithm described by \eqref{eq:PrimalDual_2},
\eqref{eq:PrimalDual_3}, \eqref{eq:PrimalDual_4} and \eqref{eq:PrimalDual-5}
is in general \textit{not implementable}. This is because \textit{neither}
the functions $g_{\mu_{S}}^{o},$ $\bbg_{\mu_{S}}$ and
$\overline{\bbf}_{\mu_{R}}^{\bbphi}$ (in general),
\textit{nor} their gradients are explicitly known apriori, \textit{for
any possible values of} $\mu_{S}>0$ and $\mu_{R}>0$ .

Nevertheless, by Lemma \ref{lem:gm_hm} from Section \ref{sec:Smoothing-via-Gaussian}
(assuming that the respective assumptions are satisfied), \textit{all}
three functions $g_{\mu_{S}}^{o},$ $\bbg_{\mu_{S}}$, $\overline{\bbf}_{\mu_{R}}^{\bbphi}$,
\textit{and} their derivatives are given by \textit{known expectation
functions}. What is more, all these expectation functions depend exclusively
on zeroth-order information, that is, on evaluations of $g_{\mu_{S}}^{o},$
$\bbg_{\mu_{S}}$ and $\overline{\bbf}_{\mu_{R}}^{\bbphi}$,
\textit{only}. Therefore, a stochastic gradient version of the algorithm
consisting by \eqref{eq:PrimalDual_2}, \eqref{eq:PrimalDual_3},
\eqref{eq:PrimalDual_4} and \eqref{eq:PrimalDual-5} may be readily
formulated by replacing all involved expectations as
\begin{flalign}
\hspace{-1pt}\hspace{-1pt}\hspace{-1pt}\hspace{-1pt}\hspace{-1pt}\bbx^{n+1} & \hspace{-1pt}\hspace{-1pt}\hspace{-0.5pt}\equiv\hspace{-0.5pt}\hspace{-1pt}\Pi_{\ccalX}\big\{\bbx^{n}\hspace{-1pt}\hspace{-1pt}+\hspace{-0.5pt}\hspace{-1pt}\boldsymbol{\gamma}_{\bbx}^{n}\circ(\Delta_{g}^{o}(\bbx^{n},\mu_{S},\boldsymbol{U}_{S}^{n+1})\boldsymbol{U}_{S}^{n+1}\hspace{-1pt}\hspace{-0.5pt}\nonumber \\
 & \;\;+\hspace{-1pt}\hspace{-1pt}\langle\boldsymbol{\Delta}_{\bbg}(\bbx^{n},\mu_{S},\boldsymbol{U}_{S}^{n+1}),\boldsymbol{\lambda}_{S}^{n}\rangle\boldsymbol{U}_{S}^{n+1}\hspace{-1pt}\hspace{-1pt}-\hspace{-1pt}\hspace{-1pt}\boldsymbol{\lambda}_{R}^{n})\hspace{-1pt}\big\},\hspace{-1pt}\hspace{-1pt}\hspace{-1pt}\hspace{-1pt}\label{eq:PrimalDual_1-1}\\
\hspace{-1pt}\hspace{-1pt}\hspace{-1pt}\hspace{-1pt}\hspace{-1pt}\bbtheta^{n+1} & \hspace{-1pt}\hspace{-1pt}\hspace{-0.5pt}\equiv\hspace{-0.5pt}\hspace{-1pt}\Pi_{\Theta}\big\{\bbtheta^{n}\hspace{-1pt}\hspace{-1pt}\nonumber \\
 & \;\;+\hspace{-1pt}\boldsymbol{\gamma}_{\bbtheta}^{n}\hspace{-1pt}\circ\hspace{-1pt}\langle\boldsymbol{\Delta}_{\bbf}(\bbtheta^{n},\mu_{R},\boldsymbol{U}_{R}^{n+1},\bbH^{n+1}),\boldsymbol{\lambda}_{R}^{n}\rangle\boldsymbol{U}_{R}^{n+1}\big\},\hspace{-1pt}\hspace{-1pt}\hspace{-1pt}\hspace{-1pt}\hspace{-1pt}\hspace{-1pt}\label{eq:PrimalDual_2-1}\\
\hspace{-1pt}\hspace{-1pt}\hspace{-1pt}\hspace{-1pt}\hspace{-1pt}\boldsymbol{\lambda}_{S}^{n+1} & \hspace{-1pt}\hspace{-1pt}\hspace{-0.5pt}\equiv\hspace{-0.5pt}\hspace{-0.5pt}\hspace{-1pt}(\boldsymbol{\lambda}_{S}^{n}\hspace{-1pt}-\hspace{-1pt}\boldsymbol{\gamma}_{\boldsymbol{\lambda}_{S}}^{n}\hspace{-1pt}\circ\hspace{0.5pt}\bbg(\bbx^{n+1}\hspace{-1pt}+\hspace{-1pt}\mu_{S}\boldsymbol{U}_{S}^{n+1}))_{+}\quad\text{and}\label{eq:PrimalDual_4-1}\\
\hspace{-1pt}\hspace{-1pt}\hspace{-1pt}\hspace{-1pt}\hspace{-1pt}\boldsymbol{\lambda}_{R}^{n+1} & \hspace{-1pt}\hspace{-1pt}\hspace{-0.5pt}\equiv\hspace{-0.5pt}\hspace{-0.5pt}\hspace{-1pt}\hspace{-0.5pt}\big(\boldsymbol{\lambda}_{R}^{n}\hspace{-1pt}-\hspace{-1pt}\boldsymbol{\gamma}_{\boldsymbol{\lambda}_{R}}^{n}\hspace{-1pt}\circ\hspace{-1pt}\big(\bbf(\bbphi(\bbH^{n+1},\bbtheta^{n+1}\hspace{-1pt}\hspace{-1pt}+\hspace{-1pt}\mu_{R}\boldsymbol{U}_{R}^{n+1}),\bbH^{n+1})\hspace{-1pt}\hspace{-1pt}\hspace{-1pt}\hspace{-1pt}\nonumber \\
 & \;\;-\hspace{-1pt}\bbx^{n+1}\hspace{-1pt}-\hspace{-1pt}\mathtt{S}(\mu_{R})\big)\hspace{-0.5pt}\big)_{+},\label{eq:PrimalDual_4-2}
\end{flalign}
where, dropping dependencies, the vectors of finite differences $\boldsymbol{\Delta}_{\bbg}\in\bbR^{N_{\bbg}}$
and $\boldsymbol{\Delta}_{\bbg}\in\bbR^{N_{S}}$ are
defined as
\begin{equation}
\boldsymbol{\Delta}_{\bbg}\triangleq\big[\Delta_{g}^{1}\,\ldots\,\Delta_{g}^{N_{\bbg}}\big]^{\boldsymbol{T}}\:\text{and}\:\boldsymbol{\Delta}_{\bbf}\triangleq\big[\Delta_{f}^{1}\,\ldots\,\Delta_{f}^{N_{S}}\big]^{\boldsymbol{T}},
\end{equation}
respectively. A complete description of the proposed model-free primal-dual
method is provided in Algorithm \ref{alg:Primal-Dual}. We may readily
observe that the algorithm requires exactly \textit{three system probes
per user, per iteration}.

\begin{figure}
\includegraphics[scale=0.467]{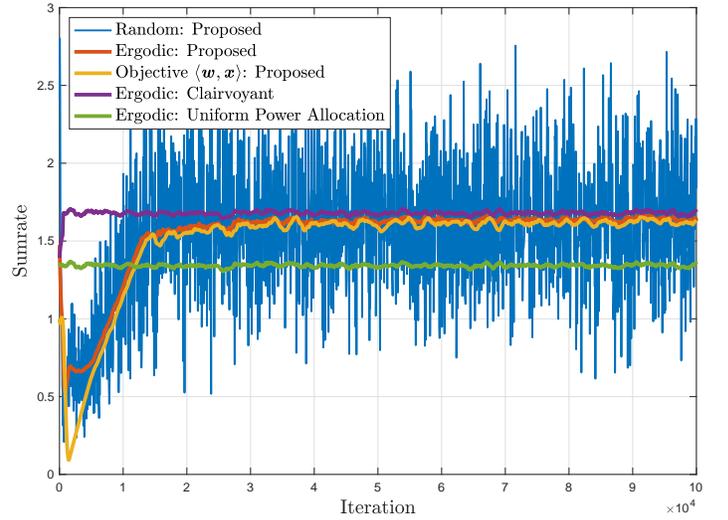}

\caption{\label{fig:AWGN_Sumrates}Sumrates achieved by the proposed method,
the clairvoyant policy, as well as a deterministic uniform power allocation
policy, in the case of a simple AWGN channel.
}
\end{figure}

Key differences between Algorithm \ref{alg:Primal-Dual} and the primal-dual
method presented in \cite{Eisen2019} are the presence of the feasibility
slack $\mathtt{S}$, which follows from our analysis, as well as the
interesting fact that, due to our explicit formulation of the smoothed
surrogate \eqref{eq:Zero_PFA}, the dual updates \eqref{eq:PrimalDual_4-1}
and \eqref{eq:PrimalDual_4-2} are naturally \textit{randomized},
in addition to the primal updates \eqref{eq:PrimalDual_1-1} and \eqref{eq:PrimalDual_2-1}. Further, while the model-free algorithm of \cite{Eisen2019} relies on policy gradient,
Algorithm \ref{alg:Primal-Dual} completely bypasses the need
for introducing randomized policies into the learning procedure. This
also makes Algorithm \ref{alg:Primal-Dual} straightforward
to implement, as computation of the gradient of
$\bbphi$ is not required; in fact, Algorithm \ref{alg:Primal-Dual}
may be executed as described for any admissible choice of $\bbphi$,
without additional computational requirements.


\begin{figure*}
\subfloat{\includegraphics[scale=0.467]{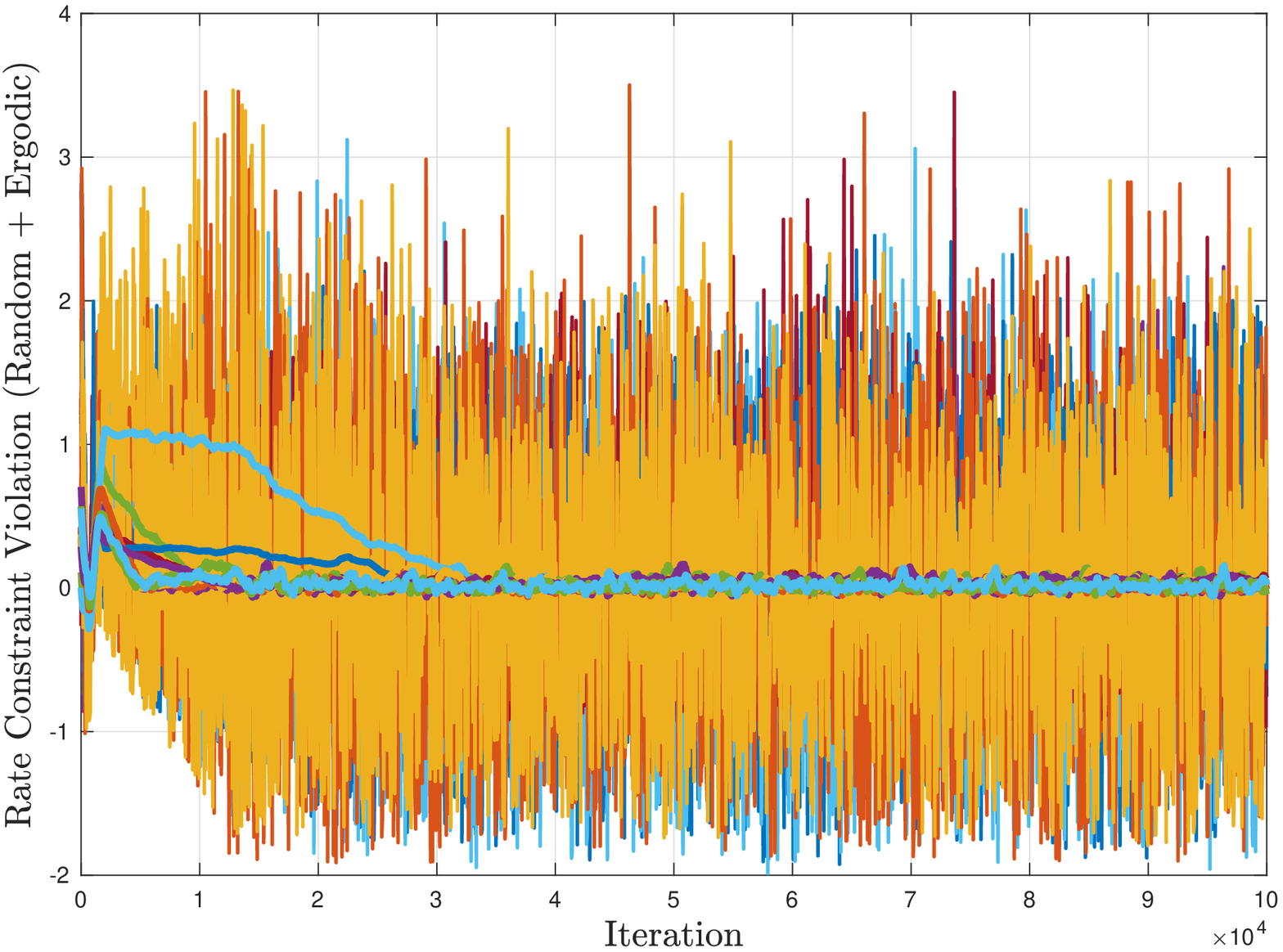}}\hspace{12.5bp}\subfloat{\includegraphics[scale=0.467]{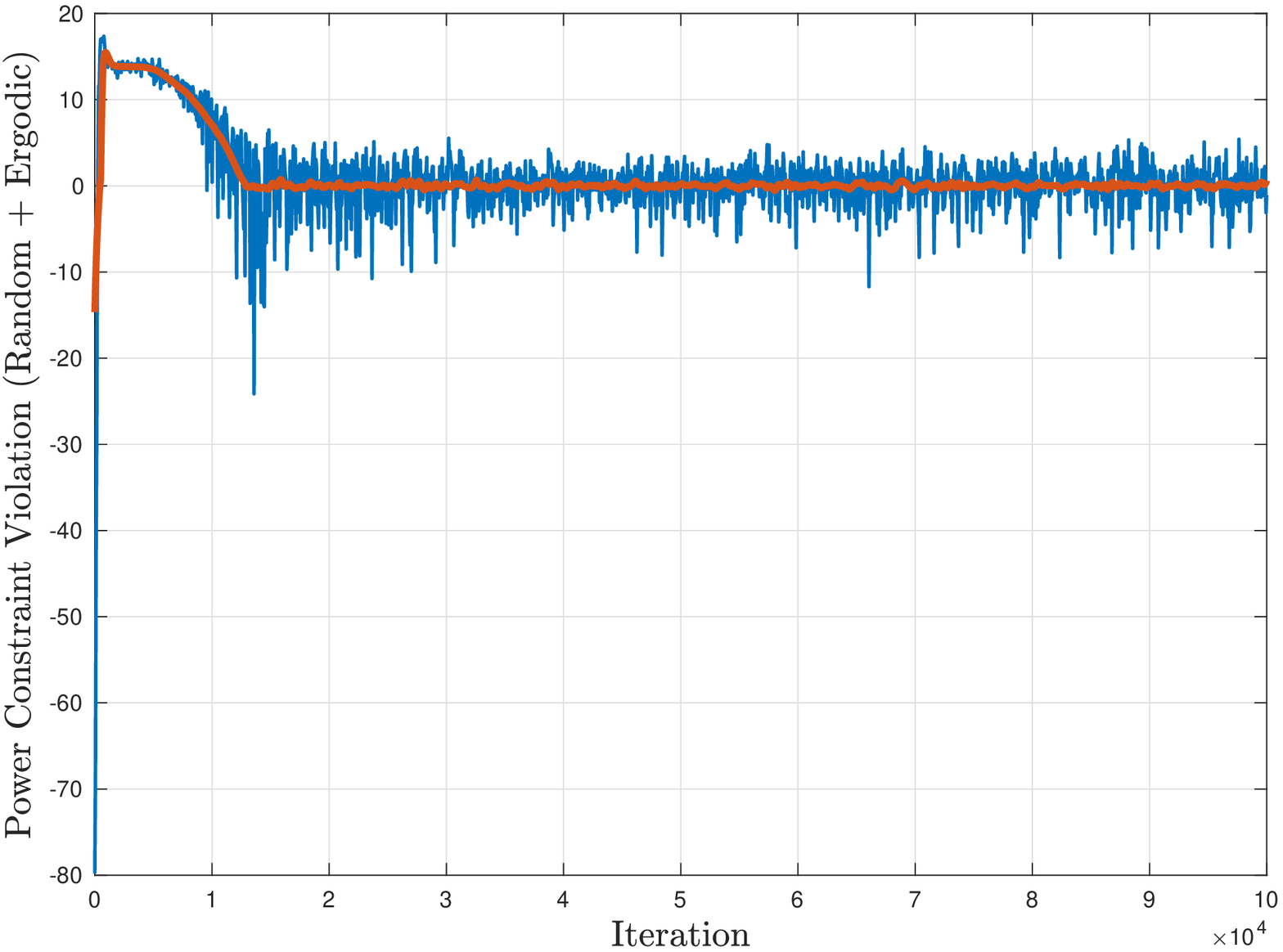}}\caption{\label{fig:AWGN_Violation}Rate (left) and power (right) constraint
violation exhibited by the proposed method, both random and ergodic,
in the case of a simple AWGN channel.}

\vspace{-14bp}
\end{figure*}

\vspace{-8bp}
\section{\label{sec:Numerical-Simulations-=000026}Numerical Simulations \&
Discussion}
We now numerically confirm and discuss the efficacy of the proposed
primal-dual algorithm (Algorithm \ref{alg:Primal-Dual}) by application
on two basic wireless models, namely, a classical Additive White Gaussian
Noise (AWGN) channel, as well as a Multiple Access Interference (MAI)
channel. Also, in all simulations presented in this section, the parameterization
$\bbphi$ is appropriately selected from the well-known
$\epsilon$-universal class of fully connected, feed-forward DNNs,
with ReLU hidden layers and sigmoid output layers, similar to the
setting considered in \cite{Eisen2019}.

For the AWGN channel case, we consider a simple multiuser networking
scenario where each user is given a dedicated channel to communicate,
with no channel interference. We wish to allocate power between users
in order to maximize the weighted sumrate of the network, within a
total expected power budget $p_{max}$, provided as a specification.
Given fixed and given user priority weights $w^{i}\ge0,i\in\mathbb{N}_{N_{S}}^{+},$
selected, without loss of generality, such that $\sum_{i}w^{i}\equiv1$,
optimal power allocation may be achieved by solving the stochastic
program
\begin{equation}
\begin{array}{rl}
\underset{x^{i},\bbtheta^{i},i\in\mathbb{N}_{N_{S}}^{+}}{\mathrm{maximize}}  & {\displaystyle \sum_{i\in\mathbb{N}_{N_{S}}^{+}}}w^{i}x^{i}\\
\mathrm{subject\,to}\hspace{1.5pt} & x^{i}\hspace{-1pt}\le\mbE\Bigg\{\hspace{-1pt}\hspace{-1pt}\log\hspace{-1pt}\hspace{-1pt}\bigg(1\hspace{-1pt}+\hspace{-1pt}\dfrac{H^{i}\phi^{i}(H^{i},\bbtheta^{i})}{\nu^{i}}\bigg)\hspace{-1pt}\hspace{-1pt}\Bigg\}\\
 & \mbE\Bigg\{{\displaystyle \sum_{i\in\mathbb{N}_{N_{S}}^{+}}}\phi^{i}(H^{i},\bbtheta^{i})\hspace{-1pt}\Bigg\}\hspace{-1pt}\le p_{max}\\
 & (x^{i},\bbtheta^{i})\in\bbR_{+}\times\bbR^{N_{\phi^{i}}},\;\forall i\in\hspace{-1pt}\mathbb{N}_{N_{S}}^{+}
\end{array}\hspace{-1pt}\hspace{-1pt}\hspace{-1pt}\hspace{-1pt},\label{eq:AWGN}
\end{equation}
where $H^{i}\ge0$ and $\nu^{i}>0$ are the fading power and noise
variance experienced by the $i$-th user, and each parameterization
$\phi^{i}:\bbR_{+}\times\bbR^{N_{\phi^{i}}}\to[0,p_{max}]$
is chosen as a DNN with single input, two hidden layers with eight
and four neurons, respectively, and a single output, for all $i\in\mathbb{N}_{N_{S}}^{+}$.
The rest of details in regard to the architecture of each of the involved
DNNs follows the discussion above. The reason for choosing $N_{S}$
uncoupled DNNs, one for each user, comes from the structure of the
globally optimal solution to the most general, unparameterized version
of problem \eqref{eq:AWGN} (mapping to \eqref{eq:Base_Problem}),
which, for this simple networking setting, may be efficiently determined
\cite{Wang2010}. Of course, this solution results in an ultimate
benchmark upper bound of the sumrate achieved by any feasible $\epsilon$-universal
resource allocation policy, at the expense of assuming complete knowledge
of the true information theoretic description of the communication
system; for this reason, we hereafter fairly refer to this solution
as \textit{clairvoyant}.

By defining vectors
\begin{flalign}
\boldsymbol{w} & \triangleq[w^{1}\,\ldots\,w^{N_{S}}]^{\boldsymbol{T}},\\
\bbphi(\bbH,\bbtheta) & \triangleq\big[\phi^{1}(H^{1},\bbtheta^{1})\,\ldots\,\phi^{N_{S}}(H^{N_{S}},\bbtheta^{N_{S}})\big]^{\boldsymbol{T}}\;\text{and}\\
\boldsymbol{\nu} & \triangleq\big[\nu^{1}\,\ldots\,\nu^{N_{S}}\big]^{\boldsymbol{T}},
\end{flalign}
problem \eqref{eq:AWGN} may be reexpressed in the canonical form
\eqref{eq:PFA} as
\begin{equation}
\hspace{-1pt}\hspace{-1pt}\hspace{-1pt}\hspace{-1pt}\hspace{-1pt}\hspace{-1pt}\hspace{-1pt}\hspace{-1pt}\hspace{-1pt}\hspace{-1pt}\hspace{-1pt}\hspace{-1pt}\begin{array}{rl}
\underset{\bbx,\bbtheta}{\mathrm{maximize}}  & \langle\boldsymbol{w},\bbx\rangle\\
\mathrm{subject\,to} & \hspace{-1pt}\begin{bmatrix}\bbx\\
0
\end{bmatrix}\hspace{-1pt}\hspace{-1pt}\le\hspace{-1pt}\mbE\hspace{-1pt}\hspace{-1pt}\begin{bmatrix}\log({\bf 1}+\bbH\circ\bbphi(\bbH,\bbtheta)\varoslash\boldsymbol{\nu})\\
p_{max}-{\bf 1}^{\boldsymbol{T}}\bbphi(\bbH,\bbtheta)
\end{bmatrix}\\
 & (\bbx,\bbtheta)\in\bbR_{+}^{N_{S}}\times\bbR^{N_{\bbphi}}
\end{array}\hspace{-1pt}\hspace{-1pt}\hspace{-1pt}\hspace{-1pt}\hspace{-1pt},\hspace{-1pt}\hspace{-1pt}\hspace{-1pt}\hspace{-1pt}\label{eq:AWGN_REF}
\end{equation}
where $\log(\cdot)$ and ``$\varoslash$'' denote the operations
of entrywise logarithm and division, respectively. Therefore, Algorithm
\ref{alg:Primal-Dual} is applicable to problem \eqref{eq:AWGN_REF}
and, in turn, \eqref{eq:AWGN}, by considering the corresponding smoothed
surrogate based on \eqref{eq:Zero_PFA}.

Since the objective of \eqref{eq:AWGN} is usually perfectly known
to the wireless engineer, we may set $\mu_{S}\equiv0$. In other words,
in the corresponding smoothed surrogate (cf. \eqref{eq:Zero_PFA}),
Gaussian smoothing is applied only to the constraints of problem \eqref{eq:AWGN_REF}.
\begin{rem}
Note that, in \eqref{eq:AWGN}, $H^{i}$ denotes the \textit{square}
of the fading channel experienced by user $i$. Nevertheless, since
the square function is one-to-one and onto on the nonnegative reals,
taking directly the square of the involved channels is virtually consistent
with our setting established in Section \ref{sec:Optimal-Resource-Allocation}.\hfill{}\ensuremath{\blacksquare}
\end{rem}
To assess the performance of the proposed Algorithm \ref{alg:Primal-Dual}
on problem \eqref{eq:AWGN_REF}, we assume $N_{S}\equiv10$ users,
set $p_{max}\equiv20$, and consider a randomly generated weight vector
$\boldsymbol{w}$. We also assume that $\nu^{i}\equiv1$, and that
$H^{i}$ is exponentially distributed with parameter $\lambda\equiv1/2$,
modeling the square of a unit variance Rayleigh fading channel state,
for all $i\in\mathbb{N}_{N_{S}}^{+}$. Then, we execute Algorithm
\ref{alg:Primal-Dual} for $10^{5}$ iterations, with initial values
$\bbx^{0}\equiv{\bf 1},\bbtheta^{0}\equiv{\bf 0}$,
$\boldsymbol{\lambda}_{S}^{0}\equiv{\bf 1}_{N_{S}+1}$, constant stepsizes
$\boldsymbol{\gamma}_{\bbx}^{n}\hspace{-1pt}\hspace{-1pt}\equiv\hspace{-1pt}\hspace{-1pt}0.001{\bf 1},\boldsymbol{\gamma}_{\bbtheta}^{n}\hspace{-1pt}\hspace{-1pt}\equiv\hspace{-1pt}\hspace{-1pt}0.0008{\bf 1},\boldsymbol{\gamma}_{\boldsymbol{\lambda}_{S}}^{n}\hspace{-1pt}\hspace{-1pt}\equiv\hspace{-1pt}\hspace{-1pt}\big[0.008{\bf 1}_{N_{S}}^{\boldsymbol{T}}\,0.0001\big]^{\boldsymbol{T}}$,
for all $n\hspace{-1pt}\hspace{-1pt}\in\hspace{-1pt}\hspace{-1pt}\mathbb{N}$,
null feasibility slack $\mathtt{S}\equiv{\bf 0}$, and with the smoothing
parameter set as $\mu_{R}\equiv10^{-9}$.

Fig. \ref{fig:AWGN_Sumrates} shows the evolution of the sequence
of objective values $\{\hspace{-0.5pt}\langle\boldsymbol{w},\bbx^{n}\rangle\hspace{-0.5pt}\}_{n\in\mathbb{N}}$,
the instantaneous sumrate sequence $\{\hspace{-0.5pt}\langle\boldsymbol{w},\log({\bf 1}+\bbH^{n}\circ\hspace{1pt}\bbphi(\bbH^{n},\bbtheta^{n})\varoslash\hspace{0.5pt}\boldsymbol{\nu})\rangle\hspace{-0.5pt}\}_{n\in\mathbb{N}}$,
as well as an approximation of the ergodic sumrate sequence $\{\hspace{-0.5pt}\langle\boldsymbol{w},\mbE\{\log({\bf 1}+\bbH\circ\bbphi(\bbH,\bbtheta^{n})\hspace{-0.5pt}\varoslash\boldsymbol{\nu})\rangle\hspace{-0.5pt}\}\hspace{-0.5pt}\}_{n\in\mathbb{N}}$.
The ergodic performance of Algorithm \ref{alg:Primal-Dual}, expressed
by the latter estimate, is also compared with the ergodic performance
of the unparameterized, globally optimal power allocation policy solving
\eqref{eq:Base_Problem} (the clairvoyant), as well as that of a deterministic
uniform power allocation policy across users. All ergodic estimates
were computed via simple moving average smoothing of the respective
process realizations.

Fig. \ref{fig:AWGN_Sumrates} readily demonstrates that the values
of the objective of \eqref{eq:AWGN_REF} match the values of the estimated
ergodic sumrate, as both obtained from Algorithm \ref{alg:Primal-Dual}.
At the same time, the ergodic sumrate obtained from Algorithm \ref{alg:Primal-Dual}
converges remarkably close to that achieved by the clairvoyant policy,
which assumes full knowledge of the model describing the wireless
system. Therefore, in this case, the proposed zeroth-order primal-dual
method attains actually near-optimal system performance.

Fig. \ref{fig:AWGN_Violation} shows similar type estimates (instantaneous
and ergodic) concerning violation of the rate and power constraints
of problem \eqref{eq:AWGN_REF}, during execution of Algorithm \ref{alg:Primal-Dual} (positive values indicate constraint violation).
We observe that all constraints are active (i.e., met with equality)
on average, which confirms that the proposed primal-dual-method indeed
converges to feasible power allocation policies, while achieving maximal
ergodic rates on a per user basis, as desired. We emphasize that,
contrary to the clairvoyant solution, such good performance of Algorithm
\ref{alg:Primal-Dual} is achieved without the availability of a baseline
model of the wireless system, and at the absence of gradient information
of information rate functions, as well as DNN parameterizations.

Next, we consider the case of a MAI channel, where\textbf{ }$N_{S}$
transmitters simultaneously communicate with a central node, for instance,
a common receiver, or a base station. In this case, the signal transmitted
by each user creates interference to the signals transmitted by all
other users in the network. As before, we would like to optimally
allocate power between users in order to maximize the weighted sumrate
of the network, within a total expected power specification $p_{max}$.
Working similarly to the AWGN channel case discussed above, we may formulate
the stochastic program
\begin{equation}
\hspace{-1pt}\hspace{-1pt}\hspace{-1pt}\hspace{-1pt}\hspace{-1pt}\hspace{-1pt}\hspace{-1pt}\hspace{-1pt}\hspace{-1pt}\hspace{-1pt}\hspace{-1pt}\hspace{-1pt}\hspace{-1pt}\hspace{-1pt}\hspace{-1pt}\begin{array}{rl}
\underset{x^{i},i\in\mathbb{N}_{N_{S}}^{+},\bbtheta}{\mathrm{maximize}}  & {\displaystyle \hspace{-1pt}\hspace{-1pt}\hspace{-1pt}\hspace{-1pt}\hspace{-1pt}\hspace{-1pt}\hspace{-1pt}\sum_{i\in\mathbb{N}_{N_{S}}^{+}}}w^{i}x^{i}\\
\mathrm{subject\,to} & \hspace{-1pt}\hspace{-1pt}\hspace{-1pt}\hspace{-1pt}\hspace{-1pt}\hspace{-1pt}\hspace{-1pt}x^{i}\hspace{-1pt}\hspace{-0.5pt}\hspace{-0.5pt}\le\hspace{-0.5pt}\mbE\Bigg\{\hspace{-1pt}\hspace{-1pt}\hspace{-0.5pt}\log\hspace{-1pt}\hspace{-1pt}\Bigg(\hspace{-1pt}1\hspace{-1pt}\hspace{-0.5pt}+\hspace{-0.5pt}\hspace{-1pt}\dfrac{H^{i}\phi^{i}(\bbH,\bbtheta)}{\nu^{i}\hspace{-1pt}+\hspace{-1pt}\sum_{i\neq j}H^{j}\phi^{j}(\bbH,\bbtheta)}\Bigg)\hspace{-1pt}\hspace{-0.5pt}\hspace{-1pt}\Bigg\}\\
 & \hspace{-1pt}\hspace{-1pt}\hspace{-1pt}\hspace{-1pt}\hspace{-1pt}\hspace{-1pt}\hspace{-1pt}\mbE\Bigg\{{\displaystyle \sum_{i\in\mathbb{N}_{N_{S}}^{+}}}\phi^{i}(\bbH,\bbtheta)\hspace{-1pt}\Bigg\}\hspace{-1pt}\hspace{-1pt}\le p_{max}\\
 & \hspace{-1pt}\hspace{-1pt}\hspace{-1pt}\hspace{-1pt}\hspace{-1pt}\hspace{-1pt}\hspace{-1pt}\hspace{-1pt}(x^{i},\bbtheta)\in\bbR_{+}\times\bbR^{N_{\bbphi}},\;\forall i\in\hspace{-1pt}\mathbb{N}_{N_{S}}^{+}
\end{array}\hspace{-1pt}\hspace{-1pt}\hspace{-1pt}\hspace{-1pt}\hspace{-0.5pt},\hspace{-1pt}\hspace{-1pt}\hspace{-1pt}\hspace{-1pt}\hspace{-1pt}\hspace{-1pt}\hspace{-1pt}\hspace{-1pt}\label{eq:DOWN_1}
\end{equation}
where each parameterization $\phi^{i}:\bbR_{+}^{N_{S}}\times\bbR^{N_{\bbphi}}\to[0,p_{max}]$,
$i\in\mathbb{N}_{N_{S}}^{+}$ is an element of the output layer of
a \textit{single} DNN taking as input the full fading channel vector
$\bbH\in\bbR_{+}^{N_{S}}$, and having two hidden
layers with thirty-two and sixteen neurons, respectively. The intuition
behind the adopted multiple-input multiple-output DNN architecture
lies in the strong coupling among the channels of all users in every
rate constraint of \eqref{eq:DOWN_1}. The rest of details in regard
to the architecture of each of the involved DNNs follows the discussion
above.

\begin{figure}
	\includegraphics[scale=0.467]{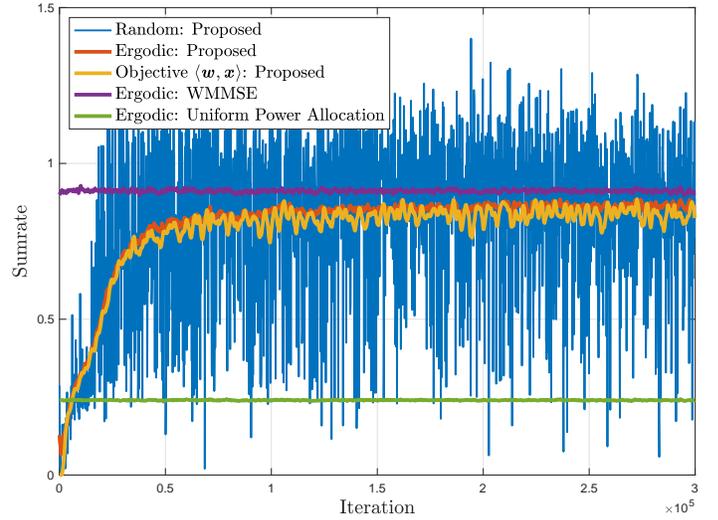}
	
	\caption{\label{fig:Down_Sumrates}Sumrates achieved by the proposed method,
		the WMMSE policy, and a deterministic uniform power allocation policy,
		in the case of a MAI channel.}
\end{figure}

As before, problem \eqref{eq:DOWN_1} may be reexpressed in the form
of \eqref{eq:PFA}; however, the details are slightly more technical
compared to the case of an AWGN channel, and are omitted for brevity.

In our simulations for this setting, we assume $N_{S}\equiv5$ users,
a power budget $p_{max}\equiv20$, and a randomly generated weight
vector $\boldsymbol{w}$. We also let $\nu^{i}\equiv1$, and $H^{i}$
follows the same exponential distribution as before, for all $i\in\mathbb{N}_{N_{S}}^{+}$.
Then, we execute Algorithm \ref{alg:Primal-Dual} for $3\cdot10^{5}$
iterations, with initial values $\bbx^{0}\equiv\bbtheta^{0}\equiv{\bf 0}$,
$\boldsymbol{\lambda}_{S}^{0}\equiv{\bf 1}_{N_{S}+1}$, constant stepsizes
$\boldsymbol{\gamma}_{\bbx}^{n}\hspace{-1pt}\hspace{-1pt}\equiv\hspace{-1pt}\hspace{-1pt}0.0008{\bf 1},\boldsymbol{\gamma}_{\bbtheta}^{n}\hspace{-1pt}\hspace{-1pt}\equiv\hspace{-1pt}\hspace{-1pt}0.0005{\bf 1},\boldsymbol{\gamma}_{\boldsymbol{\lambda}_{S}}^{n}\hspace{-1pt}\hspace{-1pt}\equiv\hspace{-1pt}\hspace{-1pt}\big[0.005{\bf 1}_{N_{S}}^{\boldsymbol{T}}\,0.0001\big]^{\boldsymbol{T}}$,
for all $n\hspace{-1pt}\hspace{-1pt}\in\hspace{-1pt}\hspace{-1pt}\mathbb{N}$,
null feasibility slack $\mathtt{S}\equiv{\bf 0}$, and with the smoothing
parameter set as $\mu_{R}\equiv10^{-9}$. As before, we set $\mu_{R}\equiv0$,
that is, the objective of \eqref{eq:DOWN_1} is reasonable assumed
known.
\begin{figure*}
\subfloat{\includegraphics[scale=0.467]{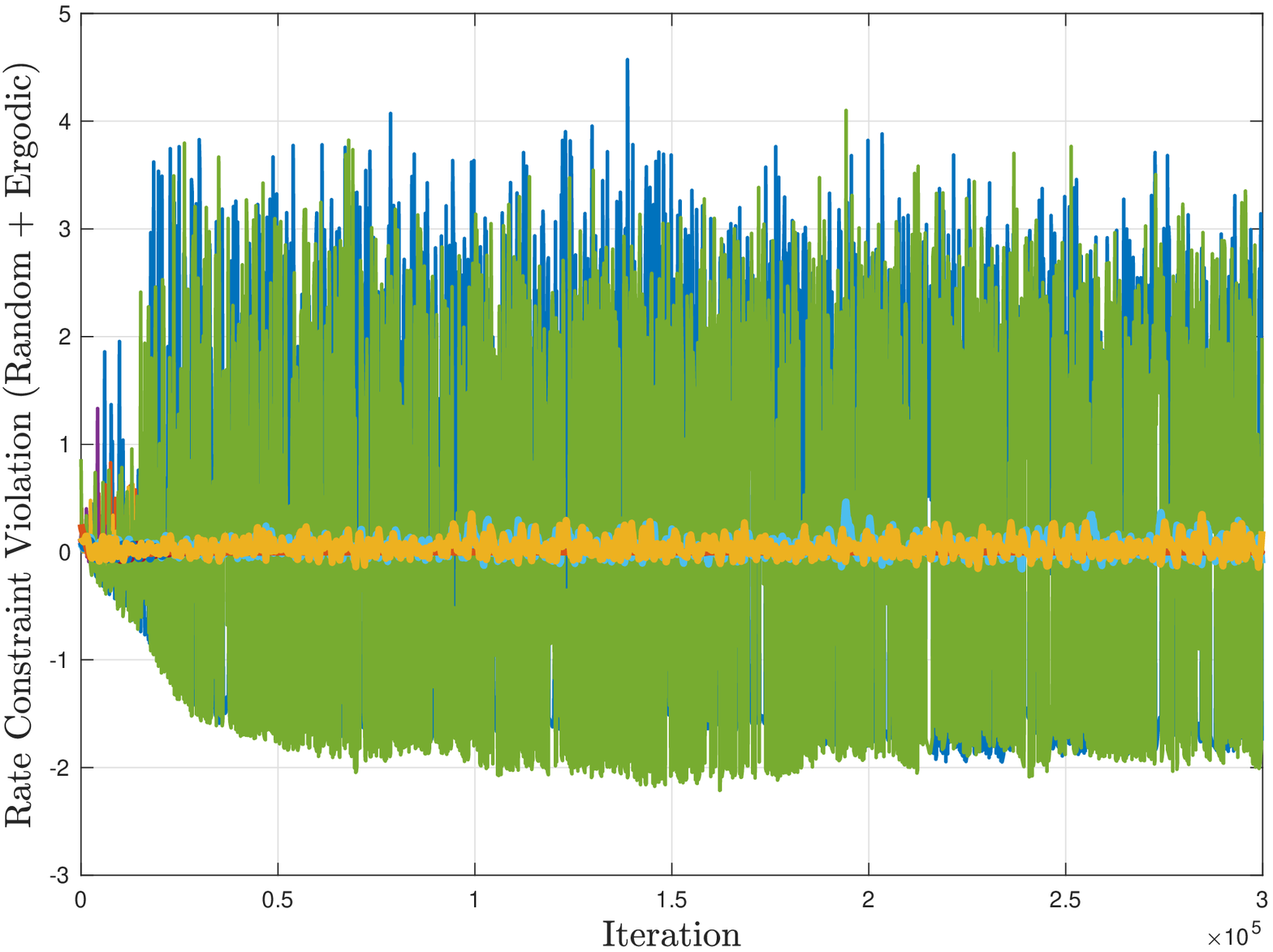}}\hspace{12.5bp}\subfloat{\includegraphics[scale=0.467]{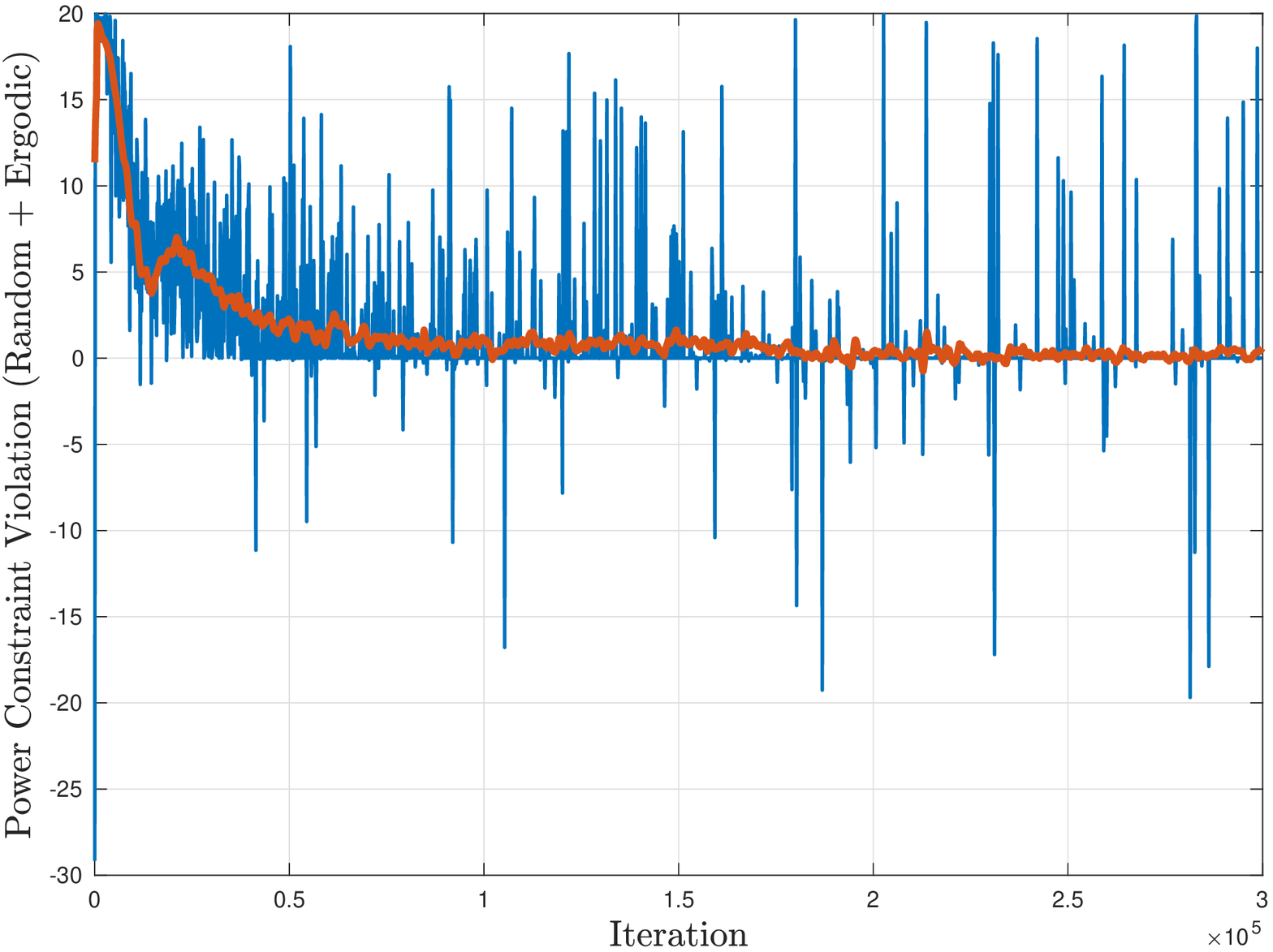}}\caption{\label{fig:DOWN_Violation}Rate (left) and power (right) constraint
violation exhibited by the proposed method, both random and ergodic,
in the case of a MAI channel.}

\vspace{-14bp}
\end{figure*}

Fig. \ref{fig:Down_Sumrates} shows the evolution of the sequence
of objective values $\{\hspace{-0.5pt}\langle\boldsymbol{w},\bbx^{n}\rangle\hspace{-0.5pt}\}_{n\in\mathbb{N}}$
and, as before, the instantaneous sumrate sequence obtained from Algorithm
\ref{alg:Primal-Dual}, as well as an approximation of the corresponding
ergodic sumrate. Since the solution to the variational version of
\eqref{eq:DOWN_1} is unavailable mainly due to nonconvexity of the
involved rate constraints (cf. \eqref{eq:Base_Problem}), we compare
the ergodic performance of Algorithm \ref{alg:Primal-Dual} with that
of the well-known WMMSE policy \cite{Shi2011}, which is an iterative
algorithm providing an approximate solution, \textit{for each fading
realization}, to the \textit{deterministic} sumrate maximization problem
\begin{equation}
\begin{array}{rl}
\underset{p^{i},i\in\mathbb{N}_{N_{S}}^{+}}{\mathrm{maximize}}  & {\displaystyle \sum_{i\in\mathbb{N}_{N_{S}}^{+}}}w^{i}\log\hspace{-1pt}\hspace{-1pt}\bigg(1\hspace{-1pt}+\hspace{-1pt}\dfrac{H^{i}p^{i}}{\nu^{i}+\hspace{-1pt}\sum_{i\neq j}H^{j}p^{j}}\bigg)\\
\mathrm{subject\,to} & {\displaystyle \sum_{i\in\mathbb{N}_{N_{S}}^{+}}}p^{i}\le p_{max}\\
 & p^{i}\ge0,\;\forall i\in\hspace{-1pt}\mathbb{N}_{N_{S}}^{+}
\end{array}\hspace{-1pt}\hspace{-1pt}\hspace{-1pt}\hspace{-1pt},\label{eq:WMMSE}
\end{equation}
Additionally, note that, as the form of problem \eqref{eq:WMMSE}
suggests, the WMMSE heuristic assumes complete knowledge of the information
theoretic model of the wireless system. For reference, Fig. \ref{fig:Down_Sumrates}
also shows the ergodic performance achieved by a uniform power allocation
policy across users. As before, all ergodic estimates were computed
via simple moving average smoothing of the respective process realizations.

Fig. \ref{fig:Down_Sumrates} confirms that Algorithm \ref{alg:Primal-Dual}
exhibits similar behavior as in the AWGN channel case previously discussed,
but for the significantly more complicated resource allocation problem
\eqref{eq:DOWN_1}. Again, the objective of \eqref{eq:DOWN_1} and
the ergodic sumrate obtained from Algorithm \ref{alg:Primal-Dual}
match, whereas the latter converges rather close to the ergodic sumrate
achieved by WMMSE.

Instantaneous and ergodic estimates of the rate and power constraint
violation of the decisions produced by Algorithm \ref{alg:Primal-Dual}
for problem \eqref{eq:DOWN_1} are provided in Fig. \ref{fig:DOWN_Violation} (again, positive values indicate constraint violation).
As in the AWGN channel case, all constraints are active on average,
confirming that the proposed zeroth-order primal-dual method produces
feasible and near-state-of-the-art power allocation policies without
knowledge of a system model and in absence of gradient information,
verifying the effectiveness of the method in a model-agnostic setting.
\section{\label{sec:Conclusion}Conclusion}
\vspace{-0bp}

We have considered the general problem of learning optimal resource
allocation policies in wireless systems, under a model-free, data-driven
setting. Starting with a generic variational formulation of the resource
allocation problem, and driven by its intractability in most wireless
networking scenarios, we focused on parametric policy function approximations.
Leveraging classical results on Gaussian smoothing, we first showed
that it is possible to crucially simplify gradient evaluation for
all utility and service functions involved, by appropriately constructing
a finite dimensional, smoothed surrogate to the original variational
problem. Then, assuming near-universal policy parameterizations,
e.g., Deep Neural Networks (DNNs), we completely characterized the
duality gap between the original problem and the dual of the proposed
surrogate, establishing linear dependence of this duality gap relative
to smoothing and near-universality parameters. In fact, this gap may
be made arbitrarily small at will. Motivated by our results, and
in conjunction with the special properties of the proposed smoothed surrogate,
we also developed a zeroth-order stochastic primal-dual algorithm,
enabling completely model-free, data-driven optimal resource allocation
for ergodic network optimization. Our simulations show that DNN-based,
data-driven policies produced by the proposed primal-dual method attain
near-ideal performance, relying exclusively on limited system probing,
completely bypassing the need for gradient computations and policy
randomization, and at the absence of baseline channel or information
rate models.

\vspace{-6bp}

\noindent \bibliographystyle{IEEEtran}
\bibliography{library_fixed}

\end{document}